\documentclass[preprint,12pt]{elsarticle}

\usepackage{amssymb}

\journal{}

\begin{document}

\begin{frontmatter}



\title{Complete classification of $(\delta+\alpha u^2)$-constacyclic codes over $\mathbb{F}_{2^m}[u]/\langle u^4\rangle$ of
oddly even length}


\author{Yonglin Cao$^{a \ \ast}$, Yuan Cao$^{b}$}

\address{$^{a}$School of Sciences,
Shandong University of Technology, Zibo, Shandong 255091, China
 \vskip 1mm $^{b}$College of Mathematics and Econometrics, Hunan University, Changsha 410082, China}
\cortext[cor1]{corresponding author. Tel: $+$86 0533 2786289; Fax: $+$86 0533 2782308. \\
E-mail addresses: ylcao@sdut.edu.cn (Yonglin Cao), \ yuan$_{-}$cao@hnu.edu.cn (Yuan Cao).}

\begin{abstract}
Let $\mathbb{F}_{2^m}$ be a finite field of cardinality $2^m$, $R=\mathbb{F}_{2^m}[u]/\langle u^4\rangle)$
and $n$ is an odd positive integer. For any $\delta,\alpha\in \mathbb{F}_{2^m}^{\times}$, ideals of the ring $R[x]/\langle x^{2n}-(\delta+\alpha u^2)\rangle$ are identified as $(\delta+\alpha u^2)$-constacyclic codes of length $2n$ over $R$. In this paper,
an explicit representation and enumeration for all distinct $(\delta+\alpha u^2)$-constacyclic codes of
length $2n$ over $R$ are presented.
\end{abstract}


\end{frontmatter}


\section{Introduction}
\noindent
  Algebraic coding theory deals with the design of error-correcting and error-detecting codes for the reliable transmission
of information across noisy channel. The class of constacyclic codes play a very significant role in
the theory of error-correcting codes.

\par
  Let $\Gamma$ be a commutative finite ring with identity $1\neq 0$, and $\Gamma^{\times}$ be the multiplicative group of invertible elements of
$\Gamma$. For any $a\in
\Gamma$, we denote by $\langle a\rangle_\Gamma$, or $\langle a\rangle$ for
simplicity, the ideal of $\Gamma$ generated by $a$, i.e., $\langle
a\rangle_\Gamma=a\Gamma=\{ab\mid b\in \Gamma\}$. For any ideal $I$ of $\Gamma$, we will identify the
element $a+I$ of the residue class ring $\Gamma/I$ with $a$ (mod $I$) for
any $a\in \Gamma$ in this paper.

\par
   A \textit{code} over $\Gamma$ of length $N$ is a nonempty subset ${\cal C}$ of $\Gamma^N=\{(a_0,a_1,\ldots$, $a_{N-1})\mid a_j\in\Gamma, \
j=0,1,\ldots,N-1\}$. The code ${\cal C}$
is said to be \textit{linear} if ${\cal C}$ is n $\Gamma$-submodule of $\Gamma^N$. All codes in this paper are assumed to be linear.
   Let $\gamma\in \Gamma^{\times}$.
A linear code
${\cal C}$ over $\Gamma$ of length $N$ is
called a $\gamma$-\textit{constacyclic code}
if $(\gamma c_{N-1},c_0,c_1,\ldots,c_{N-2})\in {\cal C}$ for all
$(c_0,c_1,\ldots,c_{N-1})\in{\cal C}$. Particularly, ${\cal C}$ is
called a \textit{negacyclic code} if $\gamma=-1$, and ${\cal C}$ is
called a  \textit{cyclic code} if $\gamma=1$.
  For any $a=(a_0,a_1,\ldots,a_{N-1})\in \Gamma^N$, let
$a(x)=a_0+a_1x+\ldots+a_{N-1}x^{N-1}\in \Gamma[x]/\langle x^N-\gamma\rangle$. We will identify $a$ with $a(x)$ in
this paper. By [9] Propositions 2.2 and 2.4, it is well known that ${\cal C}$ is a  $\gamma$-constacyclic code
of length $N$ over $\Gamma$ if and only if ${\cal C}$ is an ideal of
the residue class ring $\Gamma[x]/\langle x^N-\gamma\rangle$.

\par
  Let $\mathbb{F}_{q}$ be a finite field of cardinality $q$, where
$q$ is power of a prime, and denote $R=\mathbb{F}_{q}[u]/\langle u^e\rangle
=\mathbb{F}_{q}+u\mathbb{F}_{q}+\ldots+u^e\mathbb{F}_{q}$ ($u^e=0$) where $e\geq 2$. Then
$R$ is a finite chain ring. When $e=2$, there were a lot of literatures on linear codes, cyclic codes and
constacyclice codes of length $N$ over rings $\mathbb{F}_{p^m}[u]/\langle u^2\rangle
=\mathbb{F}_{p^m}+u\mathbb{F}_{p^m}$ for various prime $p$ and positive integers $m$ and $N$.
See [2], [4], [5],[8]--[10], [12], [14] and [17], for example.

\par
   When $e\geq 3$, for the case of $p=2$ and $m=1$ Abualrub and Siap [1] studied cyclic codes over the ring $\mathbb{Z}_2+u\mathbb{Z}_2$ and $\mathbb{Z}_2+u\mathbb{Z}_2+u^2\mathbb{Z}_2$ for arbitrary length $N$, then Al-Ashker and Hamoudeh [3] extended some of the results in [1],
and studied cyclic codes of an arbitrary length over
the ring $Z_2+uZ_2+u^2Z_2+\ldots+u^{k-1}Z_2$ ($u^k=0$) for the rank and minimal spanning of this family of codes.
For the case of $m=1$, Han  et al. [13] studied cyclic codes over $R = F_p + uF_p +\ldots+ u^{k-1}F_p$ with length $p^sn$ using discrete Fourier transform.  Singh et al. [18] studied cyclic
code over the ring $\mathbb{Z}_p[u]/\langle u^k\rangle=Z_p+uZ_p+u^2Z_p+\ldots+u^{k-1}Z_p$ for any prime integer $p$ and positive integer $N$. A set of
generators, the rank and the Hamming distance of these codes were investigated.
Kai et al. [15] investigated $(1+\lambda u)$-constacyclic codes of arbitrary length over $\mathbb{F}_p[u]/\langle u^m\rangle$, where $\lambda$ is a unit in $\mathbb{F}_p[u]/\langle u^m\rangle$, and Cao [6] generalized
these results to $(1+w\gamma)$-constacyclic codes of arbitrary length over an arbitrary finite
chain ring $R$, where $w$ is a unit of $R$ and $\gamma$ generates the unique maximal ideal of $R$.
Sobhani et al. [19] showed that the Gray image of a $(1-u^{e-1})$-constacyclic code of length
$n$ is a length $p^{m(e-1)}n$ quasi-cyclic code of index $p^{m(e-1)-1}$.

\par
   Sobhani [20] determined the structure of $(\delta+\alpha u^2)$-constacyclic codes
of length $p^k$ over $\mathbb{F}_{p^m}[u]/\langle u^3\rangle$ completely, where $\delta,\alpha\in \mathbb{F}_{p^m}^\times$, and proposed some open problems and further
researches in this area:
\textsf{characterize $(\delta+\alpha u^2)$-constacyclic codes
of length $p^k$ over the finite chain ring $\mathbb{F}_{p^m}[u]/\langle u^e\rangle$ for $e\geq 4$}.
   As a natural extension, the following problem is more worthy of study:

\noindent
\textsf{characterize $(\delta+\alpha u^2)$-constacyclic codes
of arbitrary length $N$ over the finite chain ring $\mathbb{F}_{p^m}[u]/\langle u^e\rangle$ for $e\geq 4$,
where $N=p^kn$, $k$ is a positive integer and $n\in \mathbb{Z}^{+}$ satisfying ${\rm gcd}(p,n)=1$}.

\par
   In this paper, we study the latter problem for the special case of $p=2$, $k=1$, $n$ is an odd positive integer and $e=4$.
Specifically, using linear code theory over finite chain rings we provide a new way different from the methods used in [13], [18] and [20] to give a complete classification and
an explicit enumeration for $(\delta+\alpha u^2)$-constacyclic codes of length $2n$ over the finite chain ring $\mathbb{F}_{2^m}[u]/\langle u^4\rangle$.
We will adopt the following notations.

\vskip 3mm \noindent
   {\bf Notation 1.1} Let $\delta,\alpha\in \mathbb{F}_{2^m}^{\times}$ and $n$ be an odd positive integer. We denote

\vskip 2mm \par
   $\bullet$ $R=\mathbb{F}_{2^m}[u]/\langle u^4\rangle=\mathbb{F}_{2^m}
+u\mathbb{F}_{2^m}+u^2\mathbb{F}_{2^m}+u^{3}\mathbb{F}_{2^m}$ ($u^4=0$), which is a finite chain ring of $2^{4m}$ elements.

\vskip 2mm \par
   $\bullet$ $\mathcal{A}=\mathbb{F}_{2^m}[x]/\langle(x^{2n}-\delta)^2\rangle$, which is a principal ideal ring and $|\mathcal{A}|=2^{4mn}$.

\vskip 2mm \par
   $\bullet$ $\mathcal{A}[v]/\langle v^2-\alpha^{-1}(x^{2n}-\delta)\rangle=\mathcal{A}+v\mathcal{A}$ ($v^2=\alpha^{-1}(x^{2n}-\delta)$), where
$\mathcal{A}+v\mathcal{A}=\{\xi_0+v\xi_1\mid \xi_0,\xi_1\in \mathcal{A}\}$ with operations defined by

\vskip 2mm \par
  $\diamond$ $(\xi_0+v\xi_1)+(\eta_0+v\eta_1)=(\xi_0+\eta_0)+v(\xi_1+\eta_1)$,

\vskip 2mm \par
  $\diamond$ $(\xi_0+v\xi_1)(\eta_0+v\eta_1)=(\xi_0\eta_0+\alpha^{-1}(x^{2n}-\delta)\xi_1\eta_1)+v(\xi_0\eta_1+\xi_1\eta_0)$,

\vskip 2mm \noindent
  for all $\xi_0,\xi_1,\eta_0,\eta_1\in \mathcal{A}$.

\vskip 2mm
\par
   The present paper is organized as follows.
In Section 2, we sketch the basic theory of finite commutative chain rings and linear codes over finite commutative chain rings.
In Section 3, we provide an explicit representation for each $(\delta+\alpha u^2)$-contacyclic code over $R$ of length $2n$ and give a formula to count the number of codewords in each code. As a corollary, we obtain a formula to count the number of all such codes.
Finally, we list all $258741$ distinct $(1+u^2)$-contacyclic codes of length $14$ over $\mathbb{F}_2[u]/\langle u^4\rangle$ in Section 4.



\section{Preliminaries}
\noindent
  In this section, we sketch the basic theory of finite commutative chain rings and linear codes over finite commutative chain rings needed in this paper.

\par
   Let $\mathcal{K}$ be a commutative finite chain ring with
$1\neq 0$, $\pi$ be a fixed generator of the maximal ideal of $\mathcal{K}$ with nilpotency index $4$,
and $F$ the residue field of $\mathcal{K}$
modulo its ideal $\langle \pi\rangle=\pi \mathcal{K}$, i.e.
$F=\mathcal{K}/\langle\pi \rangle$.
It is known that $|F|$ is a power of a
prime number, and
there is a unit $\xi$ of $\mathcal{K}$ with multiplicative order $|F|-1$ such
that every element $a\in \mathcal{K}$ has a unique \textit{$\pi$-adic expansion}:
$a_0+\pi a_1+\pi^2a_2+\pi^3a_3$, $a_0,a_1,a_2,a_3\in \mathcal{T}$,
where ${\cal T}=\{0,1,\xi,\ldots,\xi^{|F|-2}\}$ is the Teichm\"{u}ller
set of $\mathcal{K}$ (cf. [16]). Hence $|\mathcal{K}|=|F|^4$. If $a\neq 0$, the \textit{$\pi$-degree} of $a$ is defined as the least index $j\in\{0,1,2,3\}$ for which $a_j\neq 0$ and written for $\|a\|_\pi=j$. If $a=0$
we write $\|a\|_\pi=4$. It is clear that
$a\in \mathcal{K}^{\times}$ if and only if $a_0\neq 0$, i.e., $\|a\|_\pi=0$. Hence $|\mathcal{K}^{\times}|=(|F|-1)|F|^3$. Moreover,
we have $\mathcal{K}/\langle \pi^0\rangle=\{0\}$ and
$\mathcal{K}/\langle \pi^l\rangle=\{\sum_{i=0}^{l-1}\pi^ia_i\mid a_0,\ldots,a_{l-1}\in {\cal T}\}$ with $|\mathcal{K}/\langle \pi^l\rangle|=|F|^l$,
$1\leq l\leq 3$.

\par
   Let $L$ be a positive integer and $\mathcal{K}^L=\{(\alpha_1,\ldots,\alpha_L)\mid \alpha_1,\ldots,\alpha_L\in \mathcal{K}\}$
the free $\mathcal{K}$-module under
componentwise addition and multiplication with elements from $\mathcal{K}$. Recall that a \textit{linear code} $C$ over
$\mathcal{K}$ of length $L$ is a $\mathcal{K}$-submodule of $\mathcal{K}^L$, and $C$ is said to be \textit{nontrivial} if $C\neq \mathcal{K}^L$ and $C\neq 0$.

\par
   Let $C$ be a linear code over $\mathcal{K}$ of length $L$. By [16]
Definition 3.1, a matrix $G$ is called a \textit{generator matrix} for $C$ if the rows of $G$ span $C$ and none of them can be written as an $\mathcal{K}$-linear combination of the other rows of $G$. Furthermore, a generator matrix $G$ is
said to be \textit{in standard form} if there is a suitable permutation matrix $U$ of size $L\times L$ such that
\begin{equation}
G=\left(\begin{array}{ccccc}
\pi^0I_{k_0} & M_{0,1}     & M_{0,2}       & M_{0,3}       & M_{0,4}\cr
0            & \pi I_{k_1} & \pi M_{1,2}   & \pi M_{1,3}   &  \pi M_{1,4}\cr
0            & 0           & \pi^2 I_{k_2} & \pi^2 M_{2,3} &  \pi^2 M_{2,4}\cr
0            & 0           & 0             & \pi^3 I_{k_3} & \pi^3 M_{3,4} \end{array}\right)U
\end{equation}
where the columns are grouped into blocks of sizes $k_0,k_1,k_2,k_3, k$
with $k_i\geq 0$ and $k=L-(k_0+k_1+k_2+k_3)$. Of course, if $k_i=0$, the matrices $\pi^iI_{k_i}$ and $\pi^iM_{i,j}$ ($i<j$) are suppressed in $G$.
From [16] Proposition 3.2 and Theorem 3.5, we deduce the following.

\vskip 3mm \noindent
   {\bf Lemma 2.1} \textit{Let $C$ be a nonzero linear code of length $L$ over $\mathcal{K}$. Then $C$ has a generator matrix in standard form as in $(1)$. In this case, the number of codewords in $C$ is equal to} $|C|=|F|^{4k_0+3k_1+2k_2+k_3}=|\mathcal{T}|^{4k_0+3k_1+2k_2+k_3}$.

\vskip 3mm\par
   In particular, all distinct nontrivial linear codes of length $2$  over
$\mathcal{K}$ has been listed (cf. [7] Lemma 2.2 and Example 2.5). Moreover, we have

\vskip 3mm \noindent
   {\bf Theorem 2.2} \textit{Using the notations above, let
$\omega\in \mathcal{K}^{\times}$. Then every nontrivial linear code $C$ of length $2$ over
$\mathcal{K}$ satisfying the following condition
\begin{equation}
(\omega \pi^2 b, a)\in C, \ \forall (a,b)\in C
\end{equation}
has one and only one of the following matrices $G$ as their generator matrices}:

\vskip 2mm \par
   (I) \textit{$G=(\pi b,1)$, where $b\in (\mathcal{K}/\langle \pi^3\rangle)^{\times}$ satisfying $b^2=\omega$ $({\rm mod} \ \pi^2)$}.

\vskip 2mm \par

\vskip 2mm \par
   (II) \textit{$G=(0,\pi^3)$; $G=(\pi^{3} b,\pi^2)$ where $b\in \mathcal{K}/\langle \pi\rangle$};
     \textit{$G=(\pi^{2} b,\pi)$  where $b\in (\mathcal{K}/\langle \pi^{2}\rangle)^{\times}$ satisfying
$b^2=\omega$ $({\rm mod} \ \pi)$}.

\vskip 2mm \par
   (III) \textit{$G=\pi^kI_2$ where $I_2$ is the identity matrix of order $2$, $1\leq k\leq 3$}.

\vskip 2mm \par
   (IV) \textit{$G=\left(\begin{array}{cc}0 & 1\cr
\pi & 0\end{array}\right)$; $G=\left(\begin{array}{cc}\pi z & 1\cr
\pi^2 & 0\end{array}\right)$ where $z\in \mathcal{T}$};
  \textit{$G=\left(\begin{array}{cc}\pi z & 1\cr
\pi^3 & 0\end{array}\right)$ where $z\in \mathcal{K}/\langle \pi^2\rangle$
satisfying $z^2=\omega-\pi b$ $({\rm mod} \ \pi^2)$ for some $b\in \mathcal{K}$}.

\vskip 2mm \par
   (V) \textit{$G=\left(\begin{array}{cc}\pi^2 z & \pi\cr
\pi^{3} & 0\end{array}\right)$ where $z\in \mathcal{T}$}.

\vskip 3mm \noindent
   {\bf Proof.} See Appendix.
\hfill $\Box$



\section{Representation and classification of $(\delta+\alpha u^2)$-constacyclic codes over $R$ of length $2n$}
\noindent
In this section, we construct a specific ring isomorphism from $\mathcal{A}+v\mathcal{A}$ onto
$R[x]/\langle x^{2n}-(\delta+\alpha u^2)\rangle$. Then we obtain a one-to-one correspondence
between the set of ideals of $\mathcal{A}+v\mathcal{A}$ onto the set of ideas of $R[x]/\langle x^{2n}-(\delta+\alpha u^2)\rangle$.
Furthermore, we provide a direct sum decomposition
for each $(\delta+\alpha u^2)$-constacyclic code over $R$ of length $2n$.

\par
  Let $\xi_0+v\xi_1\in \mathcal{A}+v\mathcal{A}$ where $\xi_0,\xi_1\in \mathcal{A}$. Then
$\xi_s$ can be uniquely expressed
as $\xi_s=\xi_s(x)$ where $\xi_s(x)\in \mathbb{F}_{2^m}[x]$ satisfying ${\rm deg}(\xi_s(x))<4n$ (we will write ${\rm deg}(0)=-\infty$ for convenience) for $s=0,1$. Dividing $\xi_s(x)$ by $\alpha^{-1}(x^{2n}-\delta)$, we obtain
a unique pair $(a_0(x),a_2(x))$ of polynomials in $\mathbb{F}_{2^m}[x]$ such that
$$\xi_0=\xi_0(x)=a_0(x)+\alpha^{-1}(x^{2n}-\delta)a_2(x), \ {\rm deg}(a_j(x))<2n \ {\rm for} \ j=0,2,$$
and a unique pair $(a_1(x),a_3(x))$ of polynomials in $\mathbb{F}_{2^m}[x]$ such that
$$\xi_1=\xi_1(x)=a_1(x)+\alpha^{-1}(x^{2n}-\delta)a_3(x), \ {\rm deg}(a_j(x))<2n \ {\rm for} \ j=1,3.$$
Assume that $a_k(x)=\sum_{i=0}^{2n-1}a_{i,k}x^i$ where $a_{i,k}\in \mathbb{F}_{2^m}$ for all $i=0,1,\ldots,2n-1$ and
$k=0,1,2,3$. Then $\xi_0+v\xi_1$ can be expressed as a product of matrices:
$$\xi_0+v\xi_1=(1,x,\ldots,x^{2n-1})M\left(\begin{array}{c}1 \cr v \cr \alpha^{-1}(x^{2n}-\delta)\cr v\alpha^{-1}(x^{2n}-\delta)\end{array}\right),$$
where $M=\left(a_{i,k}\right)_{0\leq i\leq 2n-1, 0\leq k\leq 3}$ is a $2n\times 4$ matrix over $\mathbb{F}_{2^m}$. Now, we define
$\Psi(\xi_0+v\xi_1)=(1,x,\ldots,x^{2n-1})M\left(\begin{array}{c}1 \cr u \cr u^2\cr u^3\end{array}\right)
  =\sum_{i=0}^{2n-1}(\sum_{k=0}^3u^ka_{i,k})x^i$,
where $\sum_{k=0}^3u^ka_{i,k}\in R$ for all $i=0,1,\ldots,2n-1$, i.e.,
\begin{equation}
\Psi(\xi_0+v\xi_1)=a_0(x)+ua_1(x)+u^2a_2(x)+u^3a_3(x).
\end{equation}
It is clear that $\Psi$ is a bijection from
$\mathcal{A}+v\mathcal{A}$ onto $R[x]/\langle x^{2n}-(\delta+\alpha u^2)\rangle$. Then by
$v^2=\alpha^{-1}(x^{2n}-\delta)$, $(x^{2n}-\delta)^2=0$ in $\mathcal{A}+v\mathcal{A}$ and $x^{2n}-(\delta+\alpha u^2)=0$
in $R[x]/\langle x^{2n}-(\delta+\alpha u^2)\rangle$, one can easily verify the following conclustion.

\vskip 3mm \noindent
   {\bf Theorem 3.1} \textit{Using the notations above, $\Psi$ is a ring isomorphism from
$\mathcal{A}+v\mathcal{A}$ onto $R[x]/\langle x^{2n}-(\delta+\alpha u^2)\rangle$}.

\vskip 3mm\noindent
  {\bf Remark} It is clear that both $\mathcal{A}+v\mathcal{A}$ and $R[x]/\langle x^{2n}-(\delta+\alpha u^2)\rangle$ are $\mathbb{F}_{2^m}$-algebras of dimension $8n$. Specifically, we have the following:

\par
  $\bullet$ $\{1,x,\ldots,x^{4n-1},v,vx,\ldots,vx^{4n-1}\}$ is an $\mathbb{F}_{2^m}$-basis of $\mathcal{A}+v\mathcal{A}$.

\par
  $\bullet$ $\cup_{k=0}^3\{u^k,u^kx,u^kx^2,\ldots,u^kx^{2n-1}\}$  is an $\mathbb{F}_{2^m}$-basis of
$R[x]/\langle x^{2n}-(\delta+\alpha u^2)\rangle$.

\par
  $\bullet$  $\Psi$ is an $\mathbb{F}_{2^m}$-algebra isomorphism from
$\mathcal{A}+v\mathcal{A}$ onto $R[x]/\langle x^{2n}-(\delta+\alpha u^2)\rangle$ determined by:
$\Psi(x^i)=x^i$ if $0\leq i\leq 2n-1$, $\Psi(x^{2n})=\delta+\alpha u^2$ and $\Psi(v)=u.$

\vskip 3mm \par
   By Theorem 3.1, in order to determine all distinct $(\delta+\alpha u^2)$-constacyclic codes over $R$ of length $2n$,
i.e., all distinct
ideals of $R[x]/\langle x^{2n}-(\delta+\alpha u^2)\rangle$, it is sufficiency to list all distinct
ideals of $\mathcal{A}+v\mathcal{A}$.

\par
   Now, we investigate structures and properties
of rings $\mathcal{A}$ and $\mathcal{A}+v\mathcal{A}$.

\par
   Since $\delta\in \mathbb{F}_{2^m}^{\times}$ and $|\mathbb{F}_{2^m}^{\times}|=2^m-1$, there is a unique
$\delta_0\in \mathbb{F}_{2^m}^{\times}$ such that $\delta_0^2=\delta$, which implies
$x^{2n}-\delta=(x^n-\delta_0)^2$ in $\mathbb{F}_{2^m}[x]$.
As $n$ is odd,
    there are pairwise coprime monic
irreducible polynomials $f_1(x)$, $\ldots, f_r(x)$ in $\mathbb{F}_{2^m}[x]$ such that
$x^{n}-\delta_0=f_1(x)\ldots f_r(x)$
and
\begin{equation}
(x^{2n}-\delta)^\lambda=(x^n-\delta_0)^{2\lambda}=f_1(x)^{2\lambda}\ldots f_r(x)^{2\lambda}, \ \lambda=1, 2.
\end{equation}
For any integer $j$, $1\leq j\leq r$, we assume ${\rm deg}(f_j(x))=d_j$ and denote $F_j(x)=\frac{x^{n}-\delta_0}{f_j(x)}$.
Then $F_j(x)^4=\frac{(x^{2n}-\delta)^2}{f_j(x)^{4}}$ and ${\rm gcd}(F_j(x)^4,f_j(x)^{4})=1$. Hence there exist $g_j(x),h_j(x)\in \mathbb{F}_{q}[x]$ such that
$$g_j(x)F_j(x)^4+h_j(x)f_j(x)^{4}=1.$$

\par
  In the rest of this paper, we adopt the following notations.

\vskip 3mm \noindent
  {\bf Notation 3.2} Let $1\leq j\leq r$. We denote
$\mathcal{K}_j=\mathbb{F}_{2^m}[x]/\langle f_j(x)^{4}\rangle$ and set
\begin{equation}
\varepsilon_j(x)\equiv g_j(x)F_j(x)^4=1-h_j(x)f_j(x)^{4} \ ({\rm mod} \ (x^{2n}-\delta)^2).
\end{equation}

\vskip 3mm \par
 For the structure and properties of $\mathcal{K}_j$, we have the following lemma.

\vskip 3mm
\noindent
  {\bf Lemma 3.3} (cf. [7] Example 2.1) \textit{Let $1\leq j\leq r$. Then we have the following conclusions}.

\vskip 2mm\par
  (i) \textit{$\mathcal{K}_j$ is a finite chain ring, $f_j(x)$ generates the unique
maximal ideal $\langle f_j(x)\rangle=f_j(x)\mathcal{K}_j$ of $\mathcal{K}_j$, the nilpotency index of $f_j(x)$ is equal to $4$ and the residue class field of $\mathcal{K}_j$
modulo $\langle f_j(x)\rangle$ is $\mathcal{F}_j=\mathcal{K}_j/\langle f_j(x)\rangle\cong \mathbb{F}_{2^m}[x]/\langle f_j(x)\rangle$, where $\mathbb{F}_{2^m}[x]/\langle f_j(x)\rangle$ is an extension field of $\mathbb{F}_{2^m}$ with $2^{md_j}$ elements}.

\vskip 2mm\par
  (ii) \textit{Let ${\cal T}_j=\{\sum_{i=0}^{d_j-1}t_ix^i\mid t_0,t_1,\ldots,t_{d_j-1}\in \mathbb{F}_{2^m}\}$. Then ${\cal T}_j$ is equal to $\mathbb{F}_{2^m}[x]/\langle f_j(x)\rangle$ as sets, and every element $\xi$ of $\mathcal{K}_j$ has a unique $f_j(x)$-adic expansion:
$\xi=\sum_{k=0}^3f_j(x)^kb_k(x)$, $b_k(x)\in {\cal T}_j$ for all $k=0,1,2,3$.
Moreover, $\xi\in \mathcal{K}_j^{\times}$ if and only if $b_0(x)\neq 0$. Hence $|\mathcal{K}_j|=|{\cal T}_j|^4=2^{4md_j}$}.

\vskip 3mm \par
  Then from Chinese remainder theorem for commutative rings, we deduce the structure and
properties of the ring $\mathcal{A}$.

\vskip 3mm
\noindent
  {\bf Lemma 3.4} \textit{Using the notations above, we have the following}:

\par
  (i) \textit{$\varepsilon_1(x)+\ldots+\varepsilon_r(x)=1$, $\varepsilon_j(x)^2=\varepsilon_j(x)$
and $\varepsilon_j(x)\varepsilon_l(x)=0$  in the ring $\mathcal{A}$ for all $1\leq j\neq l\leq r$}.

\par
  (ii) \textit{$\mathcal{A}=\mathcal{A}_1\oplus\ldots \oplus\mathcal{A}_r$ where $\mathcal{A}_j=\mathcal{A}\varepsilon_j(x)$ with
$\varepsilon_j(x)$ as its multiplicative identity and satisfies $\mathcal{A}_j\mathcal{A}_l=\{0\}$ for all $1\leq j\neq l\leq r$}.

\par
  (iii) \textit{For any integer $j$, $1\leq j\leq r$, for any $a(x)\in \mathcal{K}_j$ we define
\begin{center}
$\varphi_j: a(x)\mapsto \varepsilon_j(x)a(x)$ $(${\rm mod} $(x^{2n}-\delta)^2)$.
\end{center}
Then $\varphi_j$ is a ring isomorphism from $\mathcal{K}_j$ onto $\mathcal{A}_j$}.

\par
  (iv) \textit{For any $a_j(x)\in \mathcal{K}_j$ for $j=1,\ldots,r$, define
\begin{center}
$\varphi(a_1(x),\ldots,a_r(x))=\sum_{j=1}^r\varphi_j(a_j(x))=\sum_{j=1}^r\varepsilon_j(x)a_j(x)$ $(${\rm mod} $(x^{2n}-\delta)^2)$.
\end{center}
Then
$\varphi$ is a ring isomorphism from $\mathcal{K}_1\times\ldots\times\mathcal{K}_r$ onto $\mathcal{A}$}.

\vskip 3mm \par
  By $\alpha^{-1}\in \mathbb{F}_{2^m}^{\times}$, there is a unique $\alpha_0\in \mathbb{F}_{2^m}^{\times}$ such that
$\alpha_0^2=\alpha^{-1}$.
In order to investigate the structure of $\mathcal{A}+v\mathcal{A}$ ($v^2=\alpha^{-1}(x^{2n}-\delta)=\alpha_0^2(x^{2n}-\delta)$),
we need the following lemma.

\vskip 3mm
\noindent
  {\bf Lemma 3.5} \textit{Let $1\leq j\leq r$ and denote
$\omega_j=\alpha_0F_j(x) \ ({\rm mod} \ f_j(x)^{4}).$
Then}

\vskip 2mm\par
  (i) \textit{$\omega_j\in \mathcal{K}_j^{\times}$ satisfying
$\alpha^{-1}(x^{2n}-\delta)=\omega_j^2f_j(x)^2 \ {\rm in} \ \mathcal{K}_j$}.

\vskip 2mm\par
  (ii) \textit{$\alpha^{-1}(x^{2n}-\delta)=\sum_{j=1}^r\varepsilon_j(x)\omega_j^2f_j(x)^2$}.

\vskip 2mm\par
  (iii) \textit{The congruence equation $z^2\equiv \omega_j^2$ $({\rm mod} \ f_j(x))$ has a unique
solution $z=\omega_j$ $({\rm mod} \ f_j(x))$}.

\vskip 2mm\par
  (iv) \textit{The congruence equation $z^2\equiv \omega_j^2$ $({\rm mod} \ f_j(x)^2)$ has $2^{md_j}$
solutions:
$$z=\omega_j+f_j(x)c(x) \ ({\rm mod} \ f_j(x)^2), \ c(x)\in \mathcal{T}_j$$
where
${\cal T}_j=\{\sum_{i=0}^{d_j-1}t_ix^i\mid t_0,t_1,\ldots,t_{d_j-1}\in \mathbb{F}_{2^m}\}$}.

\vskip 3mm
\noindent
  {\bf Proof.} (i) Since $\omega_j\in \mathcal{K}_j$ satisfying $\omega_j\equiv\alpha_0F_j(x)$ (mod $f_j(x)^4$),
by Equation (5) and $\alpha_0^2=\alpha^{-1}$, it follows that
\begin{eqnarray*}
\left(\alpha g_j(x)F_j(x)^2\right)\omega_j^2
&\equiv&\left(\alpha g_j(x)F_j(x)^2\right)\left(\alpha^{-1}F_j(x)^2\right)=1-h_j(x)f_j(x)^{4}\\
 &\equiv& 1 \ ({\rm mod} \ f_j(x)^{4}),
\end{eqnarray*}
which implies that $(\alpha g_j(x)F_j(x)^2)\omega_j^2=1$ in the ring $\mathcal{K}_j$. Hence $\omega_j\in \mathcal{K}_j^{\times}$
and $(\omega_j^2)^{-1}=\alpha g_j(x)F_j(x)^2$ (mod $f_j(x)^{4}$).
Then by Equation (4) and $F_j(x)^2=\frac{(x^{n}-\delta_0)^2}{f_j(x)^2}=\frac{x^{2n}-\delta}{f_j(x)^2}$, we have
$$\alpha^{-1}(x^{2n}-\delta)=\alpha^{-1}f_1(x)^2\ldots f_r(x)^2=\alpha^{-1}F_j(x)^2f_j(x)^2=
\omega_j^2f_j^2(x).$$

\par
  (ii) Since $\omega_j^2=\alpha_0^2F_j(x)^2=\alpha^{-1}\cdot\frac{x^{2n}-\delta}{f_j(x)^2}$ (mod $f_j(x)^4$), there exists $b_j(x)\in\mathbb{F}_{2^m}[x]$ such that
$\alpha^{-1}\cdot \frac{x^{2n}-\delta}{f_j(x)^2}=\omega_j^2+b_j(x)f_j(x)^4$. Then by Equation (5), Lemma 3.4(i) and
$F_j(x)^{4}f_j(x)^4=(x^n-\delta_0)^4=(x^{2n}-\delta)^2=0$ in $\mathcal{A}$,
we deduce that
\begin{eqnarray*}
\sum_{j=1}^r\varepsilon_j(x)\omega_j^2f_j(x)^2
&=&\sum_{j=1}^rg_j(x)F_j(x)^{4}\left(\alpha^{-1}\frac{x^{2n}-\delta}{f_j(x)^2}-b_j(x)f_j(x)^4\right)f_j(x)^2\\
  &=&\alpha^{-1}(x^{2n}-\delta)\sum_{j=1}^rg_j(x)F_j(x)^{4}\\
    &&-\sum_{j=1}^rb_j(x)g_j(x)f_j(x)^2(F_j(x)^{4}f_j(x)^4)\\
  &=&\alpha^{-1}(x^{2n}-\delta)\sum_{j=1}^r\varepsilon_j(x)\\
  &=&\alpha^{-1}(x^{2n}-\delta).
\end{eqnarray*}

\par
   (iii) We identify $\omega_j$ with $\omega_j$ (mod $f_j(x)$). Then by (i) or its proof, we have
 $\omega_j\in(\mathbb{F}_{2^m}[x]/\langle f_j(x)\rangle)^{\times}$. Let $z\in \mathbb{F}_{2^m}[x]/\langle f_j(x)\rangle$ satisfying $z^2=\omega_j^2$. Then $(z-\omega_j)^2=0$.
Since $\mathbb{F}_{2^m}[x]/\langle f_j(x)\rangle$ is a finite field of $2^m$ elements, we have
$z-\omega_j=0$, i.e., $z=\omega_j$ in $\mathbb{F}_{2^m}[x]/\langle f_j(x)\rangle$. So the congruence equation $z^2\equiv \omega_j^2$ $({\rm mod} \ f_j(x))$ has a unique
solution $z=\omega_j$ $({\rm mod} \ f_j(x))$.

\par
   (iv) We identify $\omega_j$ with $\omega_j$ (mod $f_j(x)^2$). Then by (i) or its proof, we see that
$\omega_j$ is an element of $(\mathbb{F}_{2^m}[x]/\langle f_j(x)^2\rangle)^{\times}$. Let $z\in \mathbb{F}_{2^m}[x]/\langle f_j(x)^2\rangle$ satisfying $z^2=\omega_j^2$. Then $(z-\omega_j)^2=0$. Since $\mathbb{F}_{2^m}[x]/\langle f_j(x)^2\rangle$ is a finite chain ring,
$f_j(x)$ generates its unique maximal ideal and the nilpotency index of $f_j(x)$ is equal to $2$, we deduce that $(z-\omega_j)^2=0$
is equivalent to that $z-\omega_j\in f_j(x)(\mathbb{F}_{2^m}[x]/\langle f_j(x)^2\rangle)=\{f_j(x)c(x)\mid c(x)\in {\cal T}_j\}$.
Therefore, the congruence equation $z^2\equiv \omega_j^2$ $({\rm mod} \ f_j(x)^2)$ has exactly $2^{md_j}$
solutions: $z=\omega_j+f_j(x)c(x)$ $({\rm mod} \ f_j(x)^2)$, $c(x)\in {\cal T}_j$.
\hfill $\Box$

\vskip 3mm
\par
   Then we provide the structure of $\mathcal{A}+v\mathcal{A}$ by the following lemma.

\vskip 3mm
\noindent
  {\bf Lemma 3.6} \textit{Let $1\leq j\leq r$. Using the notations in Lemma 3.5, we denote}
\begin{equation}
\mathcal{K}_j[v]/\langle v^2-\omega_j^2f_j(x)^2\rangle=\mathcal{K}_j+v\mathcal{K}_j \ (v^2=\omega_j^2f_j(x)^2),
\end{equation}
\textit{and for any $\beta_j+v\gamma_j\in \mathcal{K}_j+v_j\mathcal{K}_j$ with $\beta_j,\gamma_j\in \mathcal{K}_j$, $j=1,\ldots,r$,we define}
\begin{equation}
\Upsilon(\beta_1+v\gamma_1,\ldots,\beta_r+v\gamma_r)
=\sum_{j=1}^r\varepsilon_j(x)(\beta_j+v\gamma_j)
\end{equation}
\textit{Then
$\Upsilon$ is a ring isomorphism from $(\mathcal{K}_1+v\mathcal{K}_1)\times\ldots\times(\mathcal{K}_r+v\mathcal{K}_r)$ onto $\mathcal{A}+v\mathcal{A}$}.

\vskip 3mm\noindent
  {\bf Proof.} The ring isomorphism $\varphi:\mathcal{K}_1\times\ldots\times\mathcal{K}_r\rightarrow\mathcal{A}$
defined in Lemma 3.4(iv) can be extended to a polynomial ring isomorphism $\Upsilon_0$ from $(\mathcal{K}_1\times\ldots\times\mathcal{K}_r)[v]=\mathcal{K}_1[v]\times\ldots\times\mathcal{K}_r[v]$ onto $\mathcal{A}[v]$ in
the natural way that
\begin{eqnarray*}
&&\Upsilon_0\left(\sum_t\beta_{1,t}v^t,\ldots,\sum_t\beta_{r,t}v^t\right)\\
  &=&\sum_t\left(\sum_{j=1}^r\varphi_j(\beta_{j,t})\right)v^t
  =\sum_t\left(\sum_{j=1}^r\varepsilon_j(x)\beta_{j,t}\right)v^t
\end{eqnarray*}
$(\forall \beta_{j,t}\in \mathcal{K}_j)$. From this, by Lemma 3.4 (i) and Lemma 3.5 (ii) we deduce
\begin{eqnarray*}
&&\Upsilon_0\left(v^2-\omega_1^2f_1(x)^2, \ldots, v^2-\omega_r^2f_r(x)^2\right)\\
  &=&\left(\sum_{j=1}^r\varepsilon_j(x)\right)v^2-\sum_{j=1}^r\varepsilon_j(x)\omega_j^2f_j(x)^2
  =v^2-\alpha^{-1}(x^{2n}-\delta).
\end{eqnarray*}
Therefore, by classical ring theory we conclude that $\Upsilon_0$ induces a surjective ring homomorphism $\Upsilon$ from
\begin{center}
$\left(\mathcal{K}_1[v]/\langle v^2-\omega_1^2f_1(x)^2\rangle\right)\times\ldots\times\left(\mathcal{K}_r[v]/\langle v^2-\omega_r^2f_r(x)^2\rangle\right)$
\end{center}
onto $\mathcal{A}[v]/\langle v^2-\alpha^{-1}(x^{2n}-\delta)\rangle$ defined by (7). From this and by
\begin{eqnarray*}
&&|\left(\mathcal{K}_1[v]/\langle v^2-\omega_1^2f_1(x)^2\rangle\right)\times\ldots\times\left(\mathcal{K}_r[v]/\langle v^2-\omega_r^2f_r(x)^2\rangle\right)|\\
 &=&\prod_{j=1}^r|\mathcal{K}_j[v]/\langle v^2-\omega_j^2f_j(x)^2\rangle|=\prod_{j=1}^r|\mathcal{K}_j|^2=\prod_{j=1}^r(2^{4md_j})^2\\
 &=&2^{8m\sum_{j=1}^rd_j}=2^{8mn}=(2^{4mn})^2=|\mathcal{A}|^2\\
 &=&|\mathcal{A}[v]/\langle v^2-\alpha^{-1}(x^{2n}-\delta)\rangle|,
\end{eqnarray*}
we deduce that $\Upsilon$ is a ring isomorphism. Finally, the conclusion follows from
$\mathcal{A}[v]/\langle v^2-\alpha^{-1}(x^{2n}-\delta)\rangle=\mathcal{A}+v\mathcal{A}$ by Notation 1.1 and
$\mathcal{K}_j[v]/\langle v^2-\omega_j^2f_j(x)^2\rangle=\mathcal{K}_j+v\mathcal{K}_j$ by (6) for all $j=1,\ldots,r$.
\hfill $\Box$

\vskip 3mm \par
   In order to determine all distinct ideals of $\mathcal{A}+v\mathcal{A}$, by Lemma 3.6
it is sufficient to list all distinct ideals of the ring $\mathcal{K}_j+v\mathcal{K}_j$ ($v^2=\omega_j^2f_j(x)^2$) for
all $j=1,\ldots,r$. Since $\mathcal{K}_j$ is a subring of $\mathcal{K}_j+v\mathcal{K}_j$,
we see that $\mathcal{K}_j+v\mathcal{K}_j$ is a free $\mathcal{K}_j$-module with a basis $\{1,v\}$. Now, we define
$$\theta_j: \mathcal{K}_j^2\rightarrow \mathcal{K}_j+v\mathcal{K}_j
\ {\rm via} \ (a_0,a_1)\mapsto a_0+va_1 \ (\forall a_0,a_1\in \mathcal{K}_j).$$
Then one can easily verify that $\theta_j$ is a $\mathcal{K}_j$-module isomorphism from $\mathcal{K}_j^2$
onto $\mathcal{K}_j+v\mathcal{K}_j$. Moreover, we have the following lemma.

\vskip 3mm
\noindent
  {\bf Lemma 3.7} \textit{Using the notations above, $C$ is an ideal
of the ring $\mathcal{K}_j+v\mathcal{K}_j$ if and only if
there is a unique $\mathcal{K}_j$-submodule $S$ of $\mathcal{K}_j^2$ satisfying}
\begin{equation}
(\omega_j^2f_j(x)^2a_1,a_0)\in S, \ \forall (a_0,a_1)\in S
\end{equation}
\textit{such that $C=\theta_j(S)$}.

  \vskip 3mm \noindent
  {\bf Proof.} Let $C$ be an ideal
of $\mathcal{K}_j+v\mathcal{K}_j$. Since $\mathcal{K}_j$ is a subring
of $\mathcal{K}_j+v\mathcal{K}_j$, we see that $C$ is a
$\mathcal{K}_j$-submodule of $\mathcal{K}_j+v\mathcal{K}_j$ satisfying
$v\xi\in C$ for any $\xi\in C$. Now, let
$S=\{(a_0,a_1)\mid a_0+va_1\in C\}=\theta_j^{-1}(C)$. Then it is obvious that
$S$ is a $\mathcal{K}_j$-submodule of $\mathcal{K}_j^2$ satisfying $C=\theta_j(S)$. Moreover,
for any $(a_0,a_1)\in S$, i.e. $a_0+va_1\in C$, by $v^2=\omega_j^2f_j(x)^2$ in $\mathcal{K}_j+v\mathcal{K}_j$ it follows
that
$\omega_j^2f_j(x)^2a_1+va_0=v(a_0+va_1)\in C$. Hence
 $(\omega_j^2f_j(x)^2a_1,a_0)\in S$.

\par
  Conversely, let $C=\theta_j(S)$ and $S$ be a $\mathcal{K}_j$-submodule of $\mathcal{K}_j^2$ satisfying
Condition (8). For any $a_0+va_1\in C$ with $(a_0,a_1)\in S$ and $b_0,b_1\in \mathcal{K}_j$, by
$v^2=\omega_j^2f_j(x)^2$ in $\mathcal{K}_j+v\mathcal{K}_j$ and
$(\omega_j^2f_j(x)^2a_1,a_0)\in S$ we have
\begin{eqnarray*}
\theta_j((b_0+vb_1)(a_0+va_1))&=&\theta_j\left(b_0(a_0+vb_1)+b_1(\omega_j^2f_j(x)^2a_1+va_0)\right)\\
  &=&b_0\theta_j(a_0+va_1)+b_1\theta_j(\omega_j^2f_j(x)^2a_1+va_0)\\
  &=&b_0(a_0,a_1)+b_1(\omega_j^2f_j(x)^2a_1,a_0)\in S,
\end{eqnarray*}
which implies $(b_0+vb_1)(a_0+va_1)\in C$. Hence $C$ is an ideal of $\mathcal{K}_j+v\mathcal{K}_j$.
\hfill $\Box$

\vskip 3mm
\noindent
  {\bf Theorem 3.8} \textit{Using the notations above, let $1\leq j\leq r$. Then all distinct
ideals of $\mathcal{K}_j+v\mathcal{K}_j$ $(v^2=\omega_j^2f_j(x)^2)$ are given by one of the following cases}:

\vskip 2mm\par
  (I) \textit{$2^{2md_j}$ ideals}:

\vskip 2mm\noindent
$\bullet$  \textit{$C_j=\langle f_j(x)(\omega_j+f_j(x)c_1(x)+f_j(x)^2c_2(x))+v\rangle$ with $|C_j|=2^{4md_j}$,
where $c_1(x),c_2(x)\in \mathcal{T}_j$}.

\vskip 2mm\par
  (II) \textit{$2^{md_j+1}+1$ ideals}:

\vskip 2mm\noindent
$\bullet$   \textit{$C_j=\langle vf_j(x)^3\rangle$ with $|C_j|=2^{md_j}$};

\vskip 2mm\noindent
$\bullet$   \textit{$C_j=\langle f_j(x)^3b(x)+vf_j(x)^2\rangle$ with $|C_j|=2^{2md_j}$, where $b(x)\in \mathcal{T}_j$};

\vskip 2mm\noindent
$\bullet$   \textit{$C_j=\langle f_j(x)^2(\omega_j+f_j(x)c(x))+vf_j(x)\rangle$ with $|C_j|=2^{3md_j}$, where $c(x)\in \mathcal{T}_j$}.

\vskip 2mm\par
  (III) \textit{$5$ ideals}:

\vskip 2mm\noindent
$\bullet$  \textit{$C_j=\langle f_j(x)^k\rangle$ with $|C_j|=2^{(8-2k)md_j}$, where $0\leq k\leq 4$}.

\vskip 2mm\par
  (IV) \textit{$2^{md_j+1}+1$ ideals}:

\vskip 2mm\noindent
  $\bullet$ \textit{$C_j=\langle v,f_j(x)\rangle$ with $|C_j|=2^{7md_j}$};

\vskip 2mm\noindent
  $\bullet$  \textit{$C_j=\langle f_j(x)c(x)+v,f_j(x)^2\rangle$ with $|C_j|=2^{6md_j}$, where $c(x)\in \mathcal{T}_j$};

\vskip 2mm\noindent
$\bullet$   \textit{$C_j=\langle f_j(x)(\omega_j+f_j(x)c(x))+v,f_j(x)^3\rangle$ with $|C_j|=2^{5md_j}$, where $c(x)\in \mathcal{T}_j$}.

\vskip 2mm\par
  (V) \textit{$2^{md_j}$ ideals}:

\vskip 2mm\noindent
$\bullet$   \textit{$C_j=\langle f_j(x)^2c(x)+vf_j(x),f_j(x)^3\rangle$ with $|C_j|=2^{4md_j}$, where $c(x)\in \mathcal{T}_j$}.

\vskip 2mm\par
  \textit{Therefore, the number of ideals of $\mathcal{K}_j+v\mathcal{K}_j$ is equal to}
$$N_{(2^m,d_j,4)}=2^{2md_j}+5\cdot 2^{md_j}+7.$$

\vskip 3mm \noindent
   {\bf Proof.} By Lemma 3.3, $\mathcal{K}_j$ is a finite chain
ring, $f_j(x)$ generates its unique maximal ideal and the nilpotency index
of $f_j(x)$ is equal to $4$. From these, by Theorem 2.2 and Lemma 3.7 we deduce that all distinct ideals of
$\mathcal{K}_j+v\mathcal{K}_j$ $(v^2=\omega_j^2f_j(x)^2)$ are given by: $C_j=\theta_j(S_j)$, where $S_j$ is a
$\mathcal{K}_j$-submodules of $\mathcal{K}_j^2$ with one of the following matrices $G_j$ as its generator matrix:

\par
   (I) $G=(f_j(x) b(x),1)$, where $b(x)\in (\mathcal{K}_j/\langle f_j(x)^3\rangle)^{\times}$ satisfying $b(x)^2=\omega_j^2$ $({\rm mod} \ f_j(x)^2)$. In this case, we have $C_{j}=\theta_j(S_j)=\langle \theta_j(f_j(x)b(x),1)\rangle=\langle f_j(x)b(x)+v\rangle$.
Then by Lemma 2.1 we have $|C_{j}|=|S_j|=|\mathcal{T}_j|^{4\cdot 1}=(2^{md_j})^4=2^{4md_j}$, since $\theta_j$ is a bijection.

\par
   By Lemma 3.5(iv), we see that $b(x)\in (\mathcal{K}_j/\langle f_j(x)^3\rangle)^{\times}$ satisfying $b(x)^2=\omega_j^2$ $({\rm mod} \ f_j(x)^2)$
if and only if $b(x)=\omega_j+f_j(x)c_1(x)+f_j(x)^2c_2(x)$ with $c_1(x),c_2(x)\in \mathcal{T}_j$.
Hnece the number of ideals is equal to $|\mathcal{T}_j|^2=2^{2md_j}$.

\par
    (II) $G=(0,f_j(x)^3)$, $G=(f_j(x)^{3} b(x),f_j(x)^2)$ where $b(x)\in \mathcal{K}_j/\langle f_j(x)\rangle=\mathcal{T}_j$, and
$G=(f_j(x)^{2} b(x),f_j(x))$  where $b(x)\in (\mathcal{K}_j/\langle f_j(x)^{2}\rangle)^{\times}$ satisfying
$b(x)^2$ $=\omega_j^2$ $({\rm mod} \ f_j(x))$. In this case, an argument similar to (II) shows that $C_j$ is equal to
one of the following ideals:

\par
   $\diamondsuit$ $C_j=\langle \theta_j(0,f_j(x)^3)\rangle=\langle vf_j(x)^3\rangle$. Then $|C_{j}|=|S_j|=|\mathcal{T}_j|^{1}=2^{md_j}$ by Lemma 2.1.
\par
   $\diamondsuit$ $C_j=\langle \theta_j(f_j(x)^{3} b(x),f_j(x)^2)\rangle=\langle f_j(x)^3b(x)+vf_j(x)^2\rangle$, where $b(x)\in \mathcal{T}_j$. By Lemma 2.1 we have $|C_{j}|=|S_j|=|\mathcal{T}_j|^{2\cdot 1}=2^{2md_j}$.

\par
   $\diamondsuit$ $C_j=\langle \theta_j(f_j(x)^{2} b(x),f_j(x))\rangle=\langle f_j(x)^2b(x)+vf_j(x)\rangle$, where
$b(x)=\omega_j+f_j(x)c(x)$ and $c(x)\in \mathcal{T}_j$ by Lemma 3.5(iii).
Then $|C_{j}|=|S_j=|\mathcal{T}_j|^{3\cdot 1}=2^{3md_j}$ by Lemma 2.1.

\par
  As stated above, we see that the number of ideals is equal to $1+2|\mathcal{T}_j|=2^{md_j+1}+1$ in this case.

\par
   (III) $G_j=f_j(x)^k I_2$, where $0\leq k\leq 3$. In this case, we have $C_{j}=\langle \theta_j(f_j(x)^k,0),\theta_j(0,f_j(x)^k)\rangle=\langle f_j(x)^k,vf_j(x)^k\rangle=\langle f_j(x)^k\rangle$. Then by Lemma 2.1 we have $|C_{j}|=|S_j|=|\mathcal{T}_j|^{(4-k)\cdot 2}=2^{(8-2k)md_j}$.
If $k=4$, we have $G_j=0$ and $C_{j}=\theta_j(S_j)=0=\langle f_j(x)^4\rangle$. Obviously,
$|C_{j}|=1=2^{(8-2\cdot 4)md_j}$.

\par
  (IV) We have one of the following three subcases:

\par
  (IV-1) $G=\left(\begin{array}{cc}0 & 1\cr
f_j(x) & 0\end{array}\right)$. In this case, $C_j=\langle \theta_j(0,1), \theta_j(f_j(x),0)\rangle=\langle v,f_j(x)\rangle$, and
$|C_j|=|S_j|=|\mathcal{T}_j|^{4\cdot 1+3\cdot 1}=2^{7md_j}$ by Lemma 2.1.

\par
  (IV-2) $G=\left(\begin{array}{cc}f_j(x) c(x) & 1\cr
f_j(x)^2 & 0\end{array}\right)$, where $c(x)\in \mathcal{T}_j$.
   In this case, we have $C_j=\langle \theta_j(f_j(x)c(x),1), \theta_j(f_j(x)^2,0)\rangle=\langle f_j(x)c(x)+v,f_j(x)^2\rangle$.
Moreover, by Lemma 2.1 we have $|C_j|=|S_j|=|\mathcal{T}_j|^{4\cdot 1+2\cdot 1}=2^{6md_j}$.

\par
  (IV-3)  $G=\left(\begin{array}{cc}f_j(x) h(x) & 1\cr
f_j(x)^3 & 0\end{array}\right)$, where $h(x)\in (\mathcal{K}_j/\langle f_j(x)^2\rangle)^{\times}$
satisfying $h(x)^2=\omega_j^2-f_j(x) b(x)$ $({\rm mod} \ f_j(x)^2)$ for some $b(x)\in \mathcal{K}_j$.

\par
  In this case, we have $C_j=\langle \theta_j(f_j(x)h(x),1), \theta_j(f_j(x)^3,0)\rangle=\langle f_j(x)h(x)+v,f_j(x)^3\rangle$.
Moreover, $|C_j|=|S_j|=|\mathcal{T}_j|^{4\cdot 1+1\cdot 1}=2^{5md_j}$ by Lemma 2.1.

\par
  By the equation $h(x)^2=\omega_j^2-f_j(x) b(x)$ $({\rm mod} \ f_j(x)^2)$, we have $h(x)^2\equiv\omega_j^2$ $({\rm mod} \ f_j(x))$, which
has a unique solution $h(x)\equiv \omega_j$ $({\rm mod} \ f_j(x))$ by Lemma 3.5(iii).
Hence there exists $c(x)\in \mathcal{T}_j$ such that $h(x)=\omega_j+f_j(x)c(x)$, which
implies that $h(x)^2=\omega_j^2+f_j(x)^2c(x)^2=\omega_j^2$ in $\mathcal{K}_j/\langle f_j(x)^2\rangle$.

\par
   Conversely, for any $c(x)\in \mathcal{T}_j$ it is obvious that $h(x)=\omega_j+f_j(x)c(x)\in (\mathcal{K}_j/\langle f_j(x)^2\rangle)^{\times}$ satisfying $h(x)^2=\omega_j^2-f_j(x) b(x)$ $({\rm mod} \ f_j(x)^2)$ for $b(x)=0$.
Therefore, we conclude that $h(x)=\omega_j+f_j(x)c(x)$ where $c(x)\in \mathcal{T}_j$.

\par
  (V) $G=\left(\begin{array}{cc}f_j(x)^2 c(x) & f_j(x)\cr
f_j(x)^{3} & 0\end{array}\right)$, where $c(x)\in \mathcal{T}_j$.
In this case, we have $C_j=\langle \theta_j(f_j(x)^2c(x),f_j(x)), \theta_j(f_j(x)^3,0)\rangle=\langle f_j(x)^2c(x)+vf_j(x),f_j(x)^3\rangle$.
Moreover, by Lemma 2.1 we have $|C_j|=|S_j|=|\mathcal{T}_j|^{3\cdot 1+1\cdot 1}=2^{4md_j}$.
\hfill $\Box$

\vskip 3mm \par
   As stated above, we list all distinct
$(\delta+\alpha u^2)$-constacyclic codes over $R$ of length $2n$ by the following theorem.

\vskip 3mm
\noindent
  {\bf Corollary 3.9} \textit{Using the notations above, all distinct $(\delta+\alpha u^2)$-constacyclic codes over $R$ of length $2n$
are given by}:
$$\mathcal{C}=\Psi\left(\oplus_{j=1}^r\varepsilon_j(x)C_{j}\right)=\oplus_{j=1}^r\Psi\left(\varepsilon_j(x)C_{j}\right),$$
\textit{where $C_{j}$ is an ideal of $\mathcal{K}_j+v\mathcal{K}_j$ $(v^2=\omega_j^2f_j(x)^2)$ listed by Theorem 3.8 for all
$1\leq j\leq r$, and the number of codewords contained in $\mathcal{C}$ is equal to $|\mathcal{C}|=\prod_{j=1}^r|C_{j}|$}.

\par
   \textit{Therefore, the number of $(\delta+\alpha u^2)$-constacyclic codes over $R$ of length $2n$ is equal to $\prod_{j=1}^r
N_{(2^m,d_j,4)}$}.

\vskip 3mm
\noindent
  {\bf Proof.} By Lemma 3.6 and and Theorem 3.1, we see that $\Psi\circ\Upsilon$ is a ring
isomorphism from $(\mathcal{K}_1+v\mathcal{K}_1)\times\ldots\times (\mathcal{K}_r+v\mathcal{K}_r)$ onto
$R[x]/\langle x^{2n}-(\delta+\alpha u^2)\rangle$. From this and by Equation (7), we deduce that
$\mathcal{C}$ is an ideal of $R[x]/\langle x^{2n}-(\delta+\alpha u^2)\rangle$ if and only if for each integer $j$,
$1\leq j\leq r$, there is a unique ideal $C_{j}$ of $\mathcal{K}_j+v\mathcal{K}_j$ such that
$\mathcal{C}=(\Psi\circ\Upsilon)(C_{1}\times\ldots\times C_{r})=\Psi(\sum_{j=1}^r\varepsilon_j(x)C_{j})
  =\oplus_{j=1}^r\Psi(\varepsilon_j(x)C_{j}).$
Moreover, by Theorem 3.1, Lemma 3.6  and Theorem 3.8 it follows that
$|\mathcal{C}|=\prod_{j=1}^r|\Psi(\varepsilon_j(x)C_{j})|
=\prod_{j=1}^r|\varepsilon_j(x)C_{j}|=\prod_{j=1}^r|C_{j}|.$

\par
  Finally, for any $1\leq j\leq r$ by Theorem 3.8 we know that $\mathcal{K}_j+v\mathcal{K}_j$ has $N_{(2^m,d_j,4)}$ distinct ideals. Hence the number of $(\delta+\alpha u^2)$-constacyclic codes over $R$ of length $2n$ is equal to $\prod_{j=1}^r
N_{(2^m,d_j,4)}$.
\hfill $\Box$

\vskip 3mm\par
   In the following, we adopt the following notations:

\vskip 2mm\par
  $\bullet$ $x^i\varepsilon_j(x)f_j(x)^l=g_{i,l}^{(j)}(x)+\alpha^{-1}(x^{2n}-\delta)h_{i,l}^{(j)}(x)\in \mathcal{A}$ where
both $g_{i,l}^{(j)}(x)$ and $h_{i,l}^{(j)}(x)$ are polynomials in $\mathbb{F}_{2^m}[x]$ having degree $<2n$, for all $i=0,1,\ldots,d_j-1$ and
$l=0,1,2,3$.

\vskip 2mm\noindent
 Then by Equation (3), it follows that
\begin{equation}
\Psi(x^i\varepsilon_j(x)f_j(x)^l)=g_{i,l}^{(j)}(x)+u^2h_{i,l}^{(j)}(x)\in R[x]/\langle x^{2n}-(\delta+\alpha u^2)\rangle.
\end{equation}
By $F_j(x)^2=\frac{x^{2n}-\delta}{f_j(x)^2}$ and Equation (4), $x^i\varepsilon_j(x)f_j(x)^l=x^ig_j(x)F_j(x)^4f_j(x)^l=
(x^{2n}-\delta)x^ig_j(x)F_j(x)^2f_j(x)^{l-2}$ for all $l=2,3$, which implies
\begin{equation}
g_{i,2}^{(j)}(x)=g_{i,3}^{(j)}(x)=0 \ ({\rm mod} \ \alpha^{-1}(x^{2n}-\delta)),
\ {\rm for} \ {\rm all} \  i=0,1,\ldots,d_j-1.
\end{equation}

\vskip 2mm\par
  $\bullet$ $\varepsilon_j(x)f_j(x)^l\omega_j=p_{l}^{(j)}(x)+\alpha^{-1}(x^{2n}-\delta)q_{l}^{(j)}(x)\in \mathcal{A}$
where both $p_{l}^{(j)}(x)$ and $q_{l}^{(j)}(x)$ are polynomials in $\mathbb{F}_{2^m}[x]$ having degree $<2n$ for all $l=1,2,3$.

\vskip 2mm\noindent
 Then by Equation (3), it follows that
\begin{equation}
\Psi(\varepsilon_j(x)f_j(x)^l\omega_j)=p_{l}^{(j)}(x)+u^2q_{l}^{(j)}(x)\in R[x]/\langle x^{2n}-(\delta+\alpha u^2)\rangle.
\end{equation}
Especially, we have
\begin{equation}
p_{2}^{(j)}(x)=p_{3}^{(j)}(x)=0 \ ({\rm mod} \ \alpha^{-1}(x^{2n}-\delta)), \ {\rm for} \ {\rm all} \  i=0,1,\ldots,d_j-1.
\end{equation}

\vskip 2mm\par
  Then by Theorem 3.8, Corollary 3.9 and $\Psi(v)=u$ we deduce the following

\vskip 3mm
\noindent
  {\bf Theorem 3.10} \textit{Using the notations above, all distinct $(\delta+\alpha u^2)$-constacyclic codes of length $2n$ over $R$
are given by:
$\mathcal{C}=\mathcal{C}_1\oplus\mathcal{C}_2\oplus\ldots\oplus\mathcal{C}_r,$ where}
\textit{for each $1\leq j\leq r$, $\mathcal{C}_j$ is subcode of $\mathcal{C}$, i.e., an ideal of $R[x]/\langle x^{2n}-(\delta+\alpha u^2)\rangle$,
given by one of the following five cases}:

\vskip 2mm\par
  (I) \textit{$2^{2md_j}$ codes}:

\vskip 2mm\par
  \textit{$|\mathcal{C}_j|=2^{4md_j}$ and $\mathcal{C}_j$,
 where $c_{1i},c_{2i}\in\mathbb{F}_{2^m}$ for all $i=0,1,\ldots,d_j-1$, has an $\mathbb{F}_{2^m}$-basis}:
$x^i\xi_0, \ x^i\xi_1, \ x^i\xi_2, \ x^i\xi_3 \ ({\rm mod} \ x^{2n}-(\delta+\alpha u^2))$, $i=0,1,\ldots,d_j-1,$
\textit{where}

\vskip 2mm\noindent
  $\xi_0=p_1^{(j)}(x)+ug_{0,0}^{(j)}(x)+u^2(q_1^{(j)}(x)+\sum_{i=0}^{d_j-1}(c_{1i}h_{i,2}^{(j)}(x)+c_{2i}h_{i,3}^{(j)}(x)))+u^3h_{0,0}^{(j)}(x)$,

\vskip 2mm\noindent
  $\xi_1=ug_{0,1}^{(j)}(x)+u^2(q_2^{(j)}(x)+\sum_{i=0}^{d_j-1}c_{1i}h_{i,3}^{(j)}(x))+u^3h_{0,1}^{(j)}(x)$,

\vskip 2mm\noindent
  $\xi_2=u^2q_3^{(j)}(x)+u^3h_{0,2}^{(j)}(x)$ \textit{and}  $\xi_3=u^3h_{0,3}^{(j)}(x)$.

\vskip 2mm\par
  (II) \textit{$2^{md_j+1}+1$ codes}:

\vskip 2mm\par
  (II-1) \textit{$|\mathcal{C}_j|=2^{md_j}$ and $\mathcal{C}_j$ has an $\mathbb{F}_{2^m}$-basis:
$x^i\xi$ $({\rm mod} \ x^{2n}-(\delta+\alpha u^2))$ for $i=0,1,\ldots,d_j-1$, where $\xi=u^3h_{0,3}^{(j)}(x)$};

\vskip 2mm\par
  (II-2) \textit{$|\mathcal{C}_j|=2^{2md_j}$ and $\mathcal{C}_j$,
 where  $b_0,b_1,\ldots,b_{d_j-1}\in \mathbb{F}_{2^m}$, has an $\mathbb{F}_{2^m}$-basis}:
 $x^i\xi_2, \ x^i\xi_3  \ ({\rm mod} \ x^{2n}-(\delta+\alpha u^2)), \ i=0,1,\ldots,d_j-1,$
\textit{where $\xi_2=u^2\sum_{i=0}^{d_j-1}b_ih_{i,3}^{(j)}(x)+u^3h_{0,2}^{(j)}(x)$ and $\xi_3=u^3h_{0,3}^{(j)}(x)$};

\vskip 2mm\par
  (II-3) \textit{$|\mathcal{C}_j|=2^{3md_j}$ and $\mathcal{C}_j$, where $c_0,c_1,\ldots,c_{d_j-1}\in \mathbb{F}_{2^m}$,
has an $\mathbb{F}_{2^m}$-basis}:
$x^i\xi_1, \ x^i\xi_2, \ x^i\xi_3  \ ({\rm mod} \ x^{2n}-(\delta+\alpha u^2))$, $i=0,1,\ldots,d_j-1,$
\textit{where}

\vskip 2mm\noindent
 $\xi_1=ug_{0,1}^{(j)}(x)+u^2(q_2^{(j)}(x)+\sum_{i=0}^{d_j-1}c_ih_{i,3}^{(j)}(x))+u^3h_{0,1}^{(j)}(x)$,

\vskip 2mm\noindent
 \textit{$\xi_2=u^2q_3^{(j)}(x)+u^3h_{0,2}^{(j)}(x)$ and $\xi_3=u^3h_{0,3}^{(j)}(x)$}.

\vskip 2mm\par
  (III) \textit{$5$ codes}:

\vskip 2mm\par
  $\diamond$ \textit{$|\mathcal{C}_j|=2^{(8-2k)md_j}$ and $\mathcal{C}_j$, where $0\leq k\leq 3$, has an $\mathbb{F}_{2^m}$-basis}:

\noindent
$x^i\xi_l, \ x^i\eta_l \ ({\rm mod} \ x^{2n}-(\delta+\alpha u^2)), \ k\leq l\leq 3 \ {\rm and} \ i=0,1,\ldots,d_j-1;$

\par
  $\diamond$ \textit{$\mathcal{C}_j=\{0\}$},

\vskip 2mm\noindent
  \textit{where
   $\xi_k=g_{0,k}^{(j)}(x)+u^2h_{0,k}^{(j)}(x)$ and $\eta_k=ug_{0,k}^{(j)}(x)+u^3h_{0,k}^{(j)}(x)$ for $k=0,1$,
   $\xi_k=u^2h_{0,k}^{(j)}(x)$ and $\eta_k=u^3h_{0,k}^{(j)}(x)$ for $k=2,3$}.

\vskip 2mm\par
  (IV) \textit{$2^{md_j+1}+1$ codes}:

\vskip 2mm\par
  (IV-1) \textit{$|\mathcal{C}_j|=2^{7md_j}$ and $\mathcal{C}_j$ has an $\mathbb{F}_{2^m}$-basis}:
$x^i\eta_0, \ x^i\xi_k, \ x^i\eta_k \ ({\rm mod} \ x^{2n}-(\delta+\alpha u^2)), k=1,2,3 \ {\rm and} \ i=0,1,\ldots,d_j-1,$
\textit{where $\eta_0$, $\xi_k$ and $\eta_k$ are given by $({\rm III})$ for all $k=1,2,3$};

\vskip 2mm\par
  (IV-2) \textit{$|\mathcal{C}_j|=2^{6md_j}$ and $\mathcal{C}_j$, where $c_0,c_1,\ldots,c_{d_j-1}\in \mathbb{F}_{2^m}$, has an $\mathbb{F}_{2^m}$-basis}:
$x^i\eta_0, \ x^i\eta_1, \ x^i\xi_k, \ x^i\eta_k \ ({\rm mod} \ x^{2n}-(\delta+\alpha u^2)), k=2,3 \ {\rm and} \ i=0,1,\ldots,d_j-1,$
\textit{where}

\vskip 2mm\noindent
   $\eta_0=\sum_{i=0}^{d_j-1}c_i g_{i,1}^{(j)}(x)+ug_{0,0}^{(j)}(x)+u^2\sum_{i=0}^{d_j-1}c_i h_{i,1}^{(j)}(x)+u^3h_{0,0}^{(j)}(x)$,

\vskip 2mm\noindent
   $\eta_1=ug_{0,1}^{(j)}(x)+u^2\sum_{i=0}^{d_j-1}c_i h_{i,2}^{(j)}(x)+u^3h_{0,1}^{(j)}(x)$, \ $\eta_3=u^3h_{0,3}^{(j)}(x)$,

\vskip 2mm\noindent
    $\eta_2=u^2\sum_{i=0}^{d_j-1}c_i h_{i,3}^{(j)}(x)+u^3h_{0,2}^{(j)}(x)$,
   $\xi_2=u^2h_{0,2}^{(j)}(x)$ and $\xi_3=u^2h_{0,3}^{(j)}(x)$;

\vskip 2mm\par
  (IV-3) \textit{$|\mathcal{C}_j|=2^{5md_j}$ and $\mathcal{C}_j$, where $c_0,c_1,\ldots,c_{d_j-1}\in \mathbb{F}_{2^m}$, has an $\mathbb{F}_{2^m}$-basis}:
$x^i\eta_0, \ x^i\eta_1, \ x^i\eta_2, \ x^i\eta_3, \ x^i\xi_3 \ ({\rm mod} \ x^{2n}-(\delta+\alpha u^2)), \ i=0,1,\ldots,d_j-1,$
\textit{where}

\vskip 2mm\noindent
   $\eta_0=p_1^{(j)}(x)+ug_{0,0}^{(j)}(x)+u^2(q_1^{(j)}(x)+\sum_{i=0}^{d_j-1}c_i h_{i,2}^{(j)}(x))+u^3h_{0,0}^{(j)}(x)$,

\vskip 2mm\noindent
   $\eta_1=ug_{0,1}^{(j)}(x)+u^2(q_2^{(j)}(x)+\sum_{i=0}^{d_j-1}c_i h_{i,3}^{(j)}(x))+u^3h_{0,1}^{(j)}(x)$,

\vskip 2mm\noindent
    $\eta_2=u^2q_3^{(j)}(x)+u^3h_{0,2}^{(j)}(x)$, \ $\eta_3=u^3h_{0,3}^{(j)}(x)$ \textit{and} $\xi_3=u^2h_{0,3}^{(j)}(x)$.

\vskip 2mm\par
  (V) \textit{$2^{md_j}$ codes}:

\vskip 2mm\par
   \textit{$|\mathcal{C}_j|=2^{4md_j}$ and $\mathcal{C}_j$, where $c_0,c_1,\ldots,c_{d_j-1}\in \mathbb{F}_{2^m}$, has an $\mathbb{F}_{2^m}$-basis}:
$x^i\eta_1, \ x^i\eta_2, \ x^i\eta_3, \ x^i\xi_3 \ ({\rm mod} \ x^{2n}-(\delta+\alpha u^2)), \ i=0,1,\ldots,d_j-1,$
\textit{where}
   $\eta_1=ug_{0,1}^{(j)}(x)+u^2\sum_{i=0}^{d_j-1}c_i h_{i,2}^{(j)}(x)+u^3h_{0,1}^{(j)}(x)$,
   $\eta_2=u^2\sum_{i=0}^{d_j-1}c_i h_{i,3}^{(j)}(x)+u^3h_{0,2}^{(j)}(x)$,
   $\eta_3=u^3h_{0,3}^{(j)}(x)$ \textit{and} $\xi_3=u^2h_{0,3}^{(j)}(x)$.

\vskip 2mm\noindent
\textit{Moreover, the number of codewords contained in $\mathcal{C}$ is equal to $|\mathcal{C}|=\prod_{j=1}^r|\mathcal{C}_{j}|$}.

\vskip 3mm \noindent
  {\bf Proof.} Let $\mathcal{C}$ be a $(\delta+\alpha u^2)$-constacyclic code of length $2n$ over $R$. By Corollary 3.9 and Theorem 3.8,
$\mathcal{C}$ can be uniquely decomposed into a direct sum of subcodes: $\mathcal{C}=\mathcal{C}_1\oplus\mathcal{C}_2\oplus\ldots\oplus
\mathcal{C}_r$, where $\mathcal{C}_j=\Psi(\varepsilon_j(x)C_j)$, $1\leq j\leq r$, and $C_j$ is an ideal
of $\mathcal{K}_j+v\mathcal{K}_j$ ($v^2=\omega_j^2f_j(x)^2$) given by Theorem 3.8.

\par
  Let $C_j$ be given by case (I), i.e., $C_j=\langle f_j(x)(\omega_j+f_j(x)c_1(x)+f_j(x)^2c_2(x))$ $+v\rangle$  with $|C_j|=2^{4md_j}$, where $c_1(x),c_2(x)\in \mathcal{T}_j=\{\sum_{i=0}^{d_j-1}t_ix^i\mid
t_0,t_1,\ldots,t_{d_j-1}$ $\in \mathbb{F}_{2^m}\}$. By Lemma 3.3, Theorem 3.8 and its  proof it follows that
\begin{eqnarray*}
C_j&=&\{\xi(f_j(x)(\omega_j+f_j(x)c_1(x)+f_j(x)^2c_2(x))+v)\mid \xi\in \mathcal{K}_j\}\\
   &=&\{\sum_{k=0}^3f_j(x)^kb_k(x)\left(f_j(x)(\omega_j+f_j(x)c_1(x)+f_j(x)^2c_2(x))+v\right)\\
   &&\mid b_{k}(x)\in \mathcal{T}_j, \
    0\leq k\leq 3\}\\
   &=&\{\sum_{k=0}^2b_{k}(x)f_j(x)^{k+1}\omega_j+\sum_{k=0,1}b_{k}(x)f_j(x)^{k+2}c_1(x)+b_{0}(x)f_j(x)^{3}c_2(x)\\
   &&+v\sum_{k=0}^3f_j(x)^kb_k(x)\mid b_{k}(x)\in \mathcal{T}_j, \
    0\leq k\leq 3\}.
\end{eqnarray*}
Let $c_s(x)=\sum_{i=0}^{d_j-1}c_{si}x^i$ with $c_{si}\in\mathbb{F}_{2^m}$ for $s=1,2$. By (9)--(12) we have
\begin{eqnarray*}
\mathcal{C}_j&=&\{b_0(x)\cdot(p_1^{(j)}(x)+ug_{0,0}^{(j)}(x)+u^2(q_1^{(j)}(x)+\sum_{i=0}^{d_j-1}(c_{1i}h_{i,2}^{(j)}(x)+c_{2i}h_{i,3}^{(j)}(x)))\\
  &&+u^3h_{0,0}^{(j)}(x))+b_2(x)\cdot(u^2q_3^{(j)}(x)+u^3h_{0,2}^{(j)}(x))+b_3(x)\cdot u^3h_{0,2}^{(j)}(x)\\
  &&+b_1(x)\cdot(ug_{0,1}^{(j)}(x)+u^2(q_2^{(j)}(x)+\sum_{i=0}^{d_j-1}c_{1i}h_{i,3}^{(j)}(x))+u^3h_{0,1}^{(j)}(x))\\
  &&\mid b_0(x),b_1(x),b_2(x),b_3(x)\in \mathcal{T}_j\}\\
  &=&\{\sum_{k=0}^3\sum_{i=0}^{d_j-1}b_{ki}x^i\xi_k\mid b_{ki}\in \mathbb{F}_{2^m}, \ 0\leq k\leq 3, \ 0\leq i\leq d_j-1\}
\end{eqnarray*}
(mod $x^{2n}-(\delta+\alpha u^2)$). From this and by $|\mathcal{C}_j|=|C_j|=2^{4md_j}$, we deduce that
$\{x^i\xi_k\mid 0\leq k\leq 3, \ 0\leq i\leq d_j-1\}$ (mod $x^{2n}-(\delta+\alpha u^2)$) is an $\mathbb{F}_{2^m}$-basis
of the $(\delta+\alpha u^2)$-constacyclic code $\mathcal{C}_j$ of length $2n$ over $R$.

\par
   Similarly, one can easily verify that the conclusions hold for cases (II)--(V). Here, we omit the proof.
\hfill $\Box$



\section{An example} \label{}
\noindent
  In this section, let $R=\mathbb{F}_2[u]/\langle u^4\rangle=\mathbb{F}_2+u\mathbb{F}_2+u^2\mathbb{F}_2+u^3\mathbb{F}_2$
($u^4=0$).
we consider $(1+u^2)$-constacyclic codes over $R$ of length $14$.
   It is known that $x^7-1=x^7+1=f_1(x)f_2(x)f_3(x)$ where $f_1(x)=x+1$, $f_2(x)=x^3+x+1$ and $f_3(x)=x^3+x^2+1$ are
irreducible polynomials in $\mathbb{F}_2[x]$. Obviously, $r=3$ and $d_j={\rm deg}(f_j(x))$ satisfying $d_1=1$, $d_2=d_3=3$.
As $m=1$, by Theorem 3.8 and Corollary 3.9,
the number of $(1+u^2)$-constacyclic codes over $R$ of length $14$ is equal to
$\prod_{j=1}^3N_{(2,d_j,4)}=(2^2+5\cdot 2+7)(2^6+5\cdot 2^3+7)^2=21\cdot 111^2=258741.$

\par
   For $1\leq j\leq 3$, using the notations of Section 3, we set $F_j(x)=\frac{x^7+1}{f_j(x)}$ and
find polynomials $g_j(x),h_j(x)\in \mathbb{F}_2[x]$ such that
$g_j(x)F_j(x)^4+h_j(x)f_j(x)^4=1.$
Then
$\varepsilon_j(x)=g_j(x)F_j(x)^4 \ ({\rm mod} \ (x^{14}+1)^2)$ and
$\omega_j=F_j(x) \ ({\rm mod} \ f_j(x)^4)$.
Precisely, we have

\par
  $\varepsilon_1(x)=1+x^4+x^8+x^{12}+x^{16}+x^{20}+x^{24}$,

\par
  $\varepsilon_2(x)=1+x^4+x^8+x^{16}$, \
  $\varepsilon_3(x)=1+x^{12}+x^{20}+x^{24}$;

\par
  $\omega_1=x^3$, \
  $\omega_2=1+x+x^2+x^4$, \
  $\omega_3=1+x^2+x^3+x^4$.

\noindent
  By Equations (9) and (11), we have

\par
  $g_{0,0}^{(1)}=\sum_{l=0}^6x^{2l}$, $h_{0,0}^{(1)}=x^2+x^6+x^{10}$; \ $g_{0,2}^{(1)}=0$, $h_{0,2}^{(1)}=\sum_{l=0}^6x^{2l}$;

\par
  $g_{0,1}^{(1)}=\sum_{k=0}^{13}x^{k}$, $h_{0,1}^{(1)}=x^2+x^3+x^6+x^7+x^{10}+x^{11}$; $g_{0,3}^{(1)}=0$, $h_{0,3}^{(1)}=\sum_{k=0}^{13}x^{k}$;

\par
  $p_1^{(1)}=\sum_{k=0}^{13}x^{k}$, $q_1^{(1)}=x+x^2+x^5+x^6+x^9+x^{10}+x^{13}$;
 $q_2^{(1)}=\sum_{l=0}^6x^{2l+1}$;

\par
  $q_3^{(1)}=1+x+x^2$.

\par
  $g_{0,0}^{(2)}=1+x^2+x^4+x^8$, $h_{0,0}^{(2)}=x^2$;

\par
  $g_{0,1}^{(2)}=1+x+x^2+x^4+x^7+x^8+x^{9}+x^{11}$, $h_{0,1}^{(2)}=x^2+x^3+x^5$;

\par
  $h_{0,2}^{(2)}=1+x^2+x^4+x^8$;
 $h_{0,3}^{(2)}=1+x+x^2+x^4+x^7+x^8+x^{9}+x^{11}$;

\par
  $g_{1,0}^{(2)}=x+x^3+x^5+x^9$, $h_{1,0}^{(2)}=x^3$;

\par
  $g_{1,1}^{(2)}=x+x^2+x^3+x^5+x^8+x^9+x^{10}+x^{12}$, $h_{1,1}^{(2)}=x^3+x^4+x^6$;

\par
 $h_{1,2}^{(2)}=x+x^3+x^5+x^9$;
$h_{1,3}^{(2)}=x+x^2+x^3+x^5+x^8+x^9+x^{10}+x^{12}$;

\par
  $g_{2,0}^{(2)}=x^2+x^4+x^6+x^{10}$, $h_{2,0}^{(2)}=x^4$;

\par
  $g_{2,1}^{(2)}=x^2+x^3+x^4+x^6+x^9+x^{10}+x^{11}+x^{13}$, $h_{2,1}^{(2)}=x^4+x^5+x^7$;

\par
   $h_{2,2}^{(2)}=x^2+x^4+x^6+x^{10}$;
$h_{2,3}^{(2)}=x^2+x^3+x^4+x^6+x^9+x^{10}+x^{11}+x^{13}$;

\par
  $p_1^{(2)}=1+x+x^2+x^4+x^7+x^8+x^{9}+x^{11}$, $q_1^{(2)}=x+x^2+x^9$;

\par
  $q_2^{(2)}=1+x+x^3+x^4+x^5+x^9+x^{10}+x^{12}$,
  $q_3^{(2)}=x$.


\par
  $g_{0,0}^{(3)}=1+x^6+x^{10}+x^{12}$, $h_{0,0}^{(3)}=x^6+x^{10}$;

\par
  $g_{0,1}^{(3)}=x+x^2+x^3+x^6+x^8+x^9+x^{10}+x^{13}$, $h_{0,1}^{(3)}=1+x+x^6+x^8+x^9+x^{10}+x^{12}+x^{13}$;

\par
  $h_{0,2}^{(3)}=x^2+x^4+x^6+x^{12}$;
 $h_{0,3}^{(3)}=1+x+x^2+x^5+x^7+x^8+x^9+x^{12}$;

\par
  $g_{1,0}^{(3)}=x+x^7+x^{11}+x^{13}$, $h_{1,0}^{(3)}=x^7+x^{11}$;

\par
  $g_{1,1}^{(3)}=1+x^2+x^3+x^4+x^7+x^9+x^{10}+x^{11}$, $h_{1,1}^{(3)}=x+x^2+x^7+x^9+x^{10}+x^{11}+x^{13}$;

\par
   $h_{1,2}^{(3)}=x^3+x^5+x^7+x^{13}$;
$h_{1,3}^{(3)}=x+x^2+x^3+x^6+x^8+x^9+x^{10}+x^{13}$;

\par
  $g_{2,0}^{(3)}=1+x^2+x^8+x^{12}$, $h_{2,0}^{(3)}=1+x^8+x^{12}$;

\par
  $g_{2,1}^{(3)}=x+x^3+x^4+x^5+x^8+x^{10}+x^{11}+x^{12}$, $h_{2,1}^{(3)}=1+x^2+x^3+x^8+x^{10}+x^{11}+x^{12}$;

\par
   $h_{2,2}^{(3)}=1+x^4+x^6+x^8$;
 $h_{2,3}^{(3)}=1+x^2+x^3+x^4+x^7+x^9+x^{10}+x^{11}$;

\par
  $p_1^{(3)}=1+x^3+x^5+x^6+x^7+x^{10}+x^{12}+x^{13}$, $q_1^{(3)}=x^5+x^6+x^{10}+x^{13}$,

\par
  $q_2^{(3)}=1+x+x^5+x^6+x^7+x^9+x^{10}+x^{12}$, $q_3^{(3)}=1+x^2+x^5+x^7+x^9$.

\vskip 3mm\par
   $\diamondsuit$ By Theorem 3.10, all distinct $258741$ $(1+u^2)$-constacyclic codes over $R$ of
length $14$ are given by:
$\mathcal{C}=\mathcal{C}_1\oplus\mathcal{C}_2\oplus\mathcal{C}_3$ with $|\mathcal{C}|=|\mathcal{C}_1||\mathcal{C}_2||\mathcal{C}_3|$,
where

\noindent
$\diamond$ $\mathcal{C}_1$ is one of the $21$ codes given by Theorem 3.10 with $j=1$ and $d_1=1$;

\noindent
$\diamond$ $\mathcal{C}_2$ is one of the $111$ bcodes given by Theorem 3.10 with $j=2$ and $d_2=3$;

\noindent
$\diamond$ $\mathcal{C}_3$ is one of the $111$ codes given by Theorem 3.10 with $j=3$ and $d_3=3$.

\par
   For example, we have $8$ codes $\mathcal{C}_2$ of type (II-2): $\mathcal{C}_2^{(b_0,b_1,b_{2})}$,
$b_0,b_1,b_{2}\in \mathbb{F}_{2}$, which has an $\mathbb{F}_{2}$-basis:
 $x^i\xi_2$, $x^i\xi_3$ (mod $x^{14}-(1+u^2)$), $i=0,1,2$,
where
\begin{eqnarray*}
\xi_2&=&u^2(b_0h_{0,3}^{(2)}(x)+b_1h_{1,3}^{(2)}(x)+b_2h_{2,3}^{(2)}(x))+u^3h_{0,2}^{(2)}(x)\\
  &=&u^2(b_0(1+x+x^2+x^4+x^7+x^8+x^{9}+x^{11})\\
  &&+b_1(x+x^2+x^3+x^5+x^8+x^9+x^{10}+x^{12})\\
  &&+b_2(x^2+x^3+x^4+x^6+x^9+x^{10}+x^{11}+x^{13}))\\
  &&+u^3(1+x^2+x^4+x^8),
\end{eqnarray*}
and $\xi_3=u^3h_{0,3}^{(j)}(x)=u^3(1+x+x^2+x^4+x^7+x^8+x^{9}+x^{11})$; and
$8$ codes $\mathcal{C}_3$ of type (II-3):
$\mathcal{C}_3^{(c_0,c_1,c_2)}$, $c_0,c_1,c_{2}\in \mathbb{F}_{2}$,
which has an $\mathbb{F}_{2}$-basis: $x^i\xi_1$, $x^i\xi_2$, $x^i\xi_3$ (mod $x^{14}-(1+u^2)$), $i=0,1,2$, where
\begin{eqnarray*}
\xi_1&=&ug_{0,1}^{(3)}(x)+u^2(q_2^{(3)}(x)+c_0h_{0,3}^{(3)}(x)+c_1h_{1,3}^{(3)}(x)+c_2h_{2,3}^{(3)}(x))+u^3h_{0,1}^{(3)}(x)\\
  &=&u(x+x^2+x^3+x^6+x^8+x^9+x^{10}+x^{13})\\
  &&+u^2(1+x+x^5+x^6+x^7+x^9+x^{10}+x^{12}\\
  &&+c_0(1+x+x^2+x^5+x^7+x^8+x^9+x^{12})\\
  &&+c_1(x+x^2+x^3+x^6+x^8+x^9+x^{10}+x^{13})\\
  &&+c_2(1+x^2+x^3+x^4+x^7+x^9+x^{10}+x^{11}))\\
  &&+u^3(1+x+x^6+x^8+x^9+x^{10}+x^{12}+x^{13}),
\end{eqnarray*}
$$
\xi_2=u^2q_3^{(3)}(x)+u^3h_{0,2}^{(3)}(x)=u^2(1+x^2+x^5+x^7+x^9)+u^3(x^2+x^4+x^6+x^{12}),
$$
and $\xi_3=u^3h_{0,3}^{(3)}(x)=u^3(1+x+x^2+x^5+x^7+x^8+x^9+x^{12})$.

\par
   Let $(\mathcal{S},+)$ be an addition commutative group of order $|\mathcal{S}|$, $N$ be a positive integer and
$\mathcal{S}^N=\{(s_0,s_1,\ldots,s_{N-1})\mid s_i\in \mathcal{S}, \ i=0,1,\ldots,N-1\}$ which is a commutative group with component-wise addition.
Let $C$ be a nonempty subset of $\mathcal{S}^N$ and $\tau$ a fixed automorphism of the group $(\mathcal{S},+)$. If $C$ is a subgroup of $(\mathcal{S}^N,+)$, $C$ is called an \textit{additive code} of length $N$ over $\mathcal{S}$. In this case, the Hamming
distance $d_H(C)$ of $C$ is equal to min$\{w_H(c)\mid 0\neq c\in C\}$ and $(N, |C|,d_H(C))$ are the basic parameters of $C$.
Moreover, $C$ is said to be
\textit{$\tau$-constacyclic} if $(\tau(s_{N-1}),s_0,s_1,\ldots,s_{N-2})\in C$ for all $(s_0,s_1,\ldots,s_{N-1})\in C$. Especially,
$C$ is cyclic if $\tau$ is the identity automorphism of $(\mathcal{S},+)$.

\par
  Now, let $\mathcal{S}_1=u\mathbb{F}_2+u^2\mathbb{F}_2+u^3\mathbb{F}_2$, $\mathcal{S}_2=u^2\mathbb{F}_2+u^3\mathbb{F}_2$ and
$\mathcal{S}_3=u^3\mathbb{F}_2$, which are addition subgroups of the ring $R$ of orders $8, 4, 2$ respectively. Let $\tau$ be the
automorphism of $(\mathcal{S}_1,+)$ defined by $\tau(au+bu^2+cu^3)=au+bu^2+(a+c)u^3$ ($\forall a,b,c\in \mathbb{F}_2$). Then we have
the following results:

\par
  $\diamond$ $\mathcal{C}_2^{(0,0,0)}$ is a cyclic additive code over $\mathcal{S}_3$ with basic parameters $(14, 2^6,4)$.

\par
  $\diamond$ $\mathcal{C}_2^{(b_0,b_1,b_{2})}$ is a cyclic additive code over $\mathcal{S}_2$ with basic parameters $(14, 2^6,$ $8)$,
where $(b_0,b_1,b_{2})\in \mathbb{F}_2^3\setminus\{(0,0,0)\}$.

\par
  $\diamond$ $\mathcal{C}_3^{(c_0,c_1,c_{2})}$ is a $\tau$-constacyclic additive code over $\mathcal{S}_1$ with basic parameters $(14, 2^9,7)$,
where $(c_0,c_1,c_{2})\in \mathbb{F}_2^3$.

\par
  $\diamond$ $\mathcal{C}_2^{(b_0,b_1,b_{2})}\oplus\mathcal{C}_3^{(c_0,c_1,c_{2})}$ is a $\tau$-constacyclic additive code over $\mathcal{S}_1$ with basic parameters $(14, 2^{15},4)$,
where $(b_0,b_1,b_{2}), (c_0,c_1,c_{2})\in \mathbb{F}_2^3$.


\section{Conclusions and further research} \label{}
\noindent
For any $\delta,\alpha\mathbb{F}_{2^m}^\times$ and positive odd integer $n$, the structure and enumeration of $(\delta+\alpha u^2)$-constacyclic codes of length $2n$ over the finite chain ring $R=\mathbb{F}_{2^m}[u]/\langle u^4\rangle$ are completely
determined. The next work is to give the dual code for each of these codes and determine its self-duality precisely.  Open
problems and further researches in this area include characterizing $(\delta+\alpha u^2)$-constacyclic
codes of arbitrary length $2^kn$ over $R=\mathbb{F}_{2^m}[u]/\langle u^e\rangle$ for $k\geq 2$ and $e\geq 4$.

\vskip 3mm \noindent {\bf Acknowledgments}
 Part of this work was done when Yonglin Cao was visiting Chern Institute of Mathematics, Nankai University, Tianjin, China. Yonglin Cao would like to thank the institution for the kind hospitality. This research is
supported in part by the National Natural Science Foundation of
China (Grant Nos. 11671235, 11471255).




\begin{thebibliography}{20}

\bibitem{s1} T. Abualrub, I. Siap, Cyclic codes over the ring $\mathbb{Z}_2+u\mathbb{Z}_2$ and $\mathbb{Z}_2+u\mathbb{Z}_2+u^2\mathbb{Z}_2$,
Des. Codes Cryptogr. {\bf 42} (2007), 273--287.

\bibitem{s2} T. Abualrub, I. Siap, Constacyclic codes over $\mathbb{F}_2+u\mathbb{F}_2$,
J. Franklin Inst. {\bf 346} (2009), 520--529.

\bibitem{s3} M. Al-Ashker, M. Hamoudeh, Cyclic codes over $Z_2+uZ_2+u^2Z_2+\ldots+u^{k-1}Z_2$,
Turk. J. Math. {\bf 35}(4) (2011), 737--749.

\bibitem{s4} M. C. V. Amerra, F. R. Nemenzo, On $(1-u)$-cyclic codes over $\mathbb{F}_{p^k}+u\mathbb{F}_{p^k}$,
Appl. Math. Lett. {\bf 21} (2008), 1129--1133.

\bibitem{s5} A. Bonnecaze, P. Udaya, Cyclic codes and self-dual codes over $\mathbb{F}_2+u\mathbb{F}_2$,
IEEE Trans. Inf. Theory {\bf 45} (1999), 1250--1255.

\bibitem{s6} Y. Cao, On constacyclic codes over finite chain rings,
Finite Fields Appl. {\bf 24} (2013), 124--135.

\bibitem{s7} Y. Cao, Y. Gao, Repeate root cyclic $\mathbb{F}_q$-linear codes over
$\mathbb{F}_{q^l}$,
Finite Fields Appl. {\bf 31} (2015), 202--227.

\bibitem{s8} H. Q. Dinh, Constacyclic codes of length $2^s$ over
Galois extension rings of $\mathbb{F}_2+u\mathbb{F}_2$, IEEE Trans.
Inform. Theory {\bf 55} (2009), 1730--1740.

\bibitem{s9} H. Q. Dinh, Constacyclic codes of length $p^s$ over
$\mathbb{F}_{p^m}+u \mathbb{F}_{p^m}$, J. Algebra, {\bf 324} (2010),
940--950.

\bibitem{s10} S. T. Dougherty, P. Gaborit, M. Harada, P. Sole, Type II codes over $\mathbb{F}_2+u\mathbb{F}_2$,
IEEE Trans. Inf. Theory {\bf 45} (1999), 32--45.

\bibitem{s11}  S. T. Dougherty,  J-L. Kim,  H. Kulosman,  H. Liu: Self-dual
codes over commutative Frobenius rings, Finite Fields Appl. {\bf 16}, 14--26 (2010).

\bibitem{s12} T. A. Gulliver, M. Harada, Construction of potimal Type IV self-dual codes over $\mathbb{F}_2+u\mathbb{F}_2$,
IEEE Trans. Inf. Theory {\bf 45} (1999), 2520--2521.

\bibitem{s13} M. Han, Y. Ye, S. Zhu, C. Xu, B. Dou,
Cyclic codes over $R = F_p + uF_p +\ldots+ u^{k-1}F_p$ with length $p^sn$,
Information Sciences, {\bf 181} (2011), 926--934.

\bibitem{s14} W. C. Huffman, On the decompostion of self-dual codes over $\mathbb{F}_2+u\mathbb{F}_2$
with an automorphism of odd prime number, Finite Fields Appl. {\bf 13} (2007), 682--712.

\bibitem{s15} X. Kai, S. Zhu, P. Li, $(1+\lambda u)$-constacyclic codes over $\mathbb{F}_p[u]/\langle u^m\rangle$,
J. Franklin Inst. {\bf 347} (2010), 751--762.

\bibitem{s16} G. Norton,  A. S\u{a}l\u{a}gean-Mandache, On the structure of linear and cyclic
codes over finite chain rings, Appl. Algebra in Engrg. Comm. Comput.
{\bf 10} (2000), 489--506.

\bibitem{s17} J. F. Qian, L. N. Zhang, S. Zhu, $(1+u)$-constacyclic and cyclic codes over $\mathbb{F}_2+u\mathbb{F}_2$£¬
Appl. Math. Lett. {\bf 19} (2006), 820--823.

\bibitem{s18} A. K. Singh, P. K. Kewat, On cyclic codes over the ring $\mathbb{Z}_p[u]/\langle u^k\rangle$,
Des. Codes Cryptogr. {\bf 72} (2015), 1--13.

\bibitem{s19} R. Sobhani, M. Esmaeili, Some constacyclic and codes over $\mathbb{F}_q[u]/\langle u^{t+1}\rangle$,
IEICE Trans. Fundam. Electron. {\bf 93} (2010), 808--813.

\bibitem{s20} R. Sobhani, Complete classification of $(\delta+\alpha u^2)$-constacyclic
codes of length $p^k$ over $\mathbb{F}_{p^m}+u\mathbb{F}_{p^m}+u^2\mathbb{F}_{p^m}$,
Finite Fields Appl. {\bf 34} (2015), 123--138.

\end{thebibliography}


\vskip 5mm \noindent
{\bf Appendix: Proof of Theorem 2.2}

\vskip 3mm
   Using the notations of Section 2, by [7] Lemma 2.2 and Example 2.5 we know that the number of
linear codes over $\mathcal{K}$ of length $2$ is equal to
\begin{center}
$|F|^4+3|F|^3+5|F|^2+7|F|+9$.
\end{center}
Moreover, every nontrivial linear code $C$ over
$\mathcal{K}$ of length $2$ has one and only one of the following matrices $G$ as their generator matrices:

\vskip 2mm \par
(i) \textit{$G=(1,a)$, $a\in \mathcal{K}$}.  \  (ii) \textit{$G=(\pi^k,\pi^{k}a)$, $a\in \mathcal{K}/\langle \pi^{4-k}\rangle$, $1\leq k\leq 3$}.

\vskip 2mm \par
(iii) \textit{$G=(\pi b,1)$, $b\in \mathcal{K}/\langle \pi^{3}\rangle$}.

\vskip 2mm \par
(iv) \textit{$G=(\pi^{k+1}b,\pi^k)$, $b\in \mathcal{K}/\langle \pi^{3-k}\rangle$, $1\leq k\leq 3$}. \ (v) \textit{$G=\pi^kI_2$, $1\leq k\leq 3$}.

\vskip 2mm \par
  (vi) \textit{$G=\left(\begin{array}{cc}1 & c\cr
0 &\pi^t\end{array}\right)$,  $c\in \mathcal{K}/\langle \pi^{t}\rangle$, $1\leq t\leq 3$}.

\vskip 2mm \par
  (vii) \textit{$G=\left(\begin{array}{cc}\pi^k & \pi^kc\cr
0 &\pi^{k+t}\end{array}\right)$,  $c\in \mathcal{K}/\langle \pi^{t}\rangle$, $1\leq t\leq 3-k$, $1\leq k\leq 2$}.

\vskip 2mm \par
    (viii) \textit{$G=\left(\begin{array}{cc}c & 1\cr \pi^t & 0\end{array}\right)$, $c\in \pi(\mathcal{K}/\langle \pi^{t}\rangle)$, $1\leq t\leq 3$}.

\vskip 2mm \par
    (ix) \textit{$G=\left(\begin{array}{cc}\pi^kc & \pi^k\cr \pi^{k+t} & 0\end{array}\right)$, $c\in \pi(\mathcal{K}/\langle \pi^{t}\rangle)$,
$1\leq t\leq 3-k$, $1\leq k\leq 2$}.

\vskip 3mm \noindent
Therefore, we only need to consider the nine cases listed above:

\vskip 2mm\par
   (i) Suppose that $C$ satisfies Condition (2). Then
$(\omega \pi^2 a,1)\in C$. Since $G$ is the generator matrix of $C$, there exists $b\in \mathcal{K}$ such that $(\omega \pi^2 a,1)=b(1,a)=(b,ba)$,
i.e., $\omega \pi^2 a=b$ and $1=ba$, which implies $\omega \pi^2 a^2=1$, and we get a contradiction.

\par
   (ii) Suppose that $C$ satisfies Condition (2). Then
$(\omega \pi^{2+k}a,\pi^k)=(\omega \pi^2\cdot \pi^k a,\pi^k)\in C$. So there exists $b\in \mathcal{K}$ such that $(\omega \pi^{2+k}a,\pi^k)=b(\pi^k,\pi^k a)=(\pi^k b,\pi^k ab)$,
which implies $\omega \pi^{2+k}a=\pi^k b$ and $\pi^k =\pi^k ba$.
Hence $\pi^k=\omega \pi^{2+k}a\cdot a=\omega \pi^{2+k}a^2$ and we get a contradiction.

\par
   (iii) In this case, $C$ satisfies Condition (2) if and only if
there exists $a\in \mathcal{K}$ such that $(\omega \pi^2,\pi b)=a(\pi b,1)=(\pi ba,a)$, i.e., $\omega \pi^2=\pi ba$ and $\pi b=a$. These conditions
are equivalent to that $b$ satisfies $\omega \pi^2=\pi b\cdot \pi b=\pi^2b$, i.e., $b^2=\omega$ (mod $\pi^2$). From this and by
$\omega\in \mathcal{K}^{\times}$, we deduce that $b\in (\mathcal{K}/\langle \pi^3\rangle)^{\times}$.

\par
   (iv) In this case, $C$ satisfies Condition (2) if and only if
there exists $a\in \mathcal{K}$ such that $(\omega \pi^{2+k},\pi^{k+1} b)=(\omega \pi^2\cdot \pi^k,\pi^{k+1} b)=a(\pi^{k+1} b,\pi^k)=(\pi^{k+1} ba,\pi^ka)$, i.e., $\omega \pi^{2+k}=\pi^{k+1} ba$ and $\pi^{k+1} b=\pi^ka$. These conditions
are equivalent to that $b$ satisfies $\omega \pi^{2+k}=\pi b\cdot \pi^k a=\pi b\cdot \pi^{k+1} b=\pi^{k+2} b^2$. Then we have the
following two subcases:

\par
  (iv-1) When $k=2,3$, we have $\pi^{2+k}=0$. Hence $C$ satisfies Condition (2) for any $b\in \mathcal{K}/\langle \pi^{3-k}\rangle$.

\par
  (iv-2) When $k=1$, then $C$ satisfies Condition (2) if and only if $b\in \mathcal{K}/\langle \pi^{2}\rangle$ satisfying
$\omega \pi^{3}=\pi^{3} b^2$, i.e., $b^2=\omega$ (mod $\pi$).

\par
   (v) In this case, $C$ satisfies Condition (2) for all $1\leq k\leq 3$.

\par
   (vi) Suppose
that $C$ satisfies Condition (2). Then there exist $a,b\in \mathcal{K}$ such that $(\omega \pi^2 c,1)=a(1,c)+b(0,\pi^t)=(a,ac+\pi^t b)$,
i.e., $\omega \pi^2 c=a$ and $1=ac+\pi^t b$, which implies $1=\omega \pi^2 c^2+\pi^t b=\pi(\omega \pi c^2+\pi^{t-1}b)$, and we get a contradiction.
Hence $C$ does not satisfy Condition (1) in this case.

\par
   (vii) Suppose
that $C$ satisfies Condition (2). Then there exist $a,b\in \mathcal{K}$ such that $(\omega \pi^2\cdot\pi^k c,\pi^k)=a(\pi^k,\pi^kc)+b(0,\pi^{k+t})
=(\pi^ka,\pi^kac+\pi^{k+t} b)$,
i.e., $\omega \pi^{k+2} c=\pi^ka$ and $\pi^k=\pi^kac+\pi^{k+t} b$, which implies $\pi^k=\omega \pi^{k+2} c^2+\pi^{k+t} b
=\pi^{k+1}(\omega \pi c^2+\pi^{t-1}b)$, and we get a contradiction.

\par
   (viii) It is clear that $(\omega\pi^2\cdot  0,\pi^t)=\pi^t(c,1)-c(\pi^t,0)\in C$.
Hence $C$ satisfies Condition (2) if and only if there exist $a,b\in \mathcal{K}$ such that
$(\omega\pi^2,c)=a(c,1)+b(\pi^t,0)=(ac+\pi^tb,a)$,
i.e., $\omega\pi^2=ac+\pi^tb$ and $c=a$, which are equivalent to
that $\omega\pi^2=c^2+\pi^tb$, i.e., $c^2=\omega\pi^2-\pi^tb$ for some $b\in \mathcal{K}$.
Then we have one of the following three subcases:

\par
   (viii-1) When $t=1$, then $c\in \pi(\mathcal{K}/\langle \pi\rangle)=\{0\}$, i.e., $c=0$ (for $b=\omega\pi$).

\par
   (viii-2) When $t=2$, then $c\in \pi(A/\langle \pi\rangle)$. It is clear that $c=0$ satisfies the
condition $c^2=\pi(\omega\pi-\pi^{2-1}b)=\pi^2(\omega-b)$ for $b=0$. Now, let $c\neq 0$. Then $1\leq\|c\|_{\pi}\leq t-1=1$, which implies $\|c\|_{\pi}=1$. Hence
$c=\pi z$ for some $z\in (\mathcal{K}/\langle \pi\rangle)^{\times}=\mathcal{T}\setminus\{0\}$.
From this and by $c^2=\omega\pi^2-\pi^tb$, we deduce that $\pi^2 z^2=\pi^2(\omega-b)$,
i.e., $z^2=\omega-b$ (mod $\pi^2$).

\par
   Let $z\in \mathcal{T}\setminus\{0\}$ arbitrary. We select $b=\omega-z^2\in \mathcal{K}$.
Then it is clear that $z^2=\omega-b$ (mod $\pi^2$).

\par
   (viii-3) When $t=3$, then $c\in \pi(\mathcal{K}/\langle \pi\rangle)$. Suppose that $c=0$. Then by $c^2=\omega\pi^2-\pi^tb$, we have that
$\omega\pi^2=\pi^3b$, which is a contradiction. Hence $c\neq 0$ satisfying $1\leq\|c\|_{\pi}\leq t-1=2$. Suppose that
$\|c\|_{\pi}\geq 2$. Then $c^2=0$, which implies that $\omega\pi^2=\pi^3b$ as well, that is impossible. Hence
$\|c\|_{\pi}=1$. So $c=\pi z$ for some $z\in (\mathcal{K}/\langle \pi^2\rangle)^{\times}$.
From this and by $c^2=\omega\pi^2-\pi^tb$, we deduce that $\pi^2 z^2=\pi^2(\omega-\pi b)$,
i.e., $z^2=\omega-\pi b$ (mod $\pi^2$).

\par
   (ix) It is clear that $(\omega\pi^2\cdot  0,\pi^{k+t})=\pi^t(\pi^kc,\pi^k)-c(\pi^{k+t},0)\in C$.
Hence $C$ satisfies Condition (2) if and only if there exist $a,b\in \mathcal{K}$ such that
$(\omega\pi^2\cdot \pi^k,\pi^kc)=a(\pi^kc,\pi^k)+b(\pi^{k+t},0)=(\pi^kac+\pi^{k+t}b,\pi^ka)$,
i.e., $\omega\pi^{k+2}=\pi^kac+\pi^{k+t}b$ and $\pi^kc=\pi^ka$, which are equivalent to
that $\omega\pi^{k+2}=\pi^kc^2+\pi^{k+t}b$, i.e., $\pi^kc^2=\omega\pi^{k+2}-\pi^{k+t}b=\pi^{k+1}(\omega\pi-\pi^{t-1}b)\in \pi^{k+1} \mathcal{K}$ for some $b\in \mathcal{K}$.

\par
  By $c\in \pi(\mathcal{K}/\langle \pi^t\rangle)$, we have $c=0$ or $c\neq0$ satisfying $1\leq\|c\|_{\pi}\leq t-1$.
The latter condition implies that $2\leq t\leq 3-k$. Hence $k=1$ and $t=2$, which implies that $\|c\|_{\pi}=1$.
Hence $c=\pi z$ where $z\in (\mathcal{K}/\langle \pi^{t-1}\rangle)^{\times}=(\mathcal{K}/\langle \pi\rangle)^{\times}=\mathcal{T}\setminus\{0\}$.
Therefore,
the condition $\omega\pi^{k+2}=\pi^kc^2+\pi^{k+t}b$
is reduced to $\pi^{3}z^2=\omega\pi^{3}-\pi^{3}b=\pi^{3}(\omega-b)$, which is equivalent to
that $z^2=\omega-b$ (mod $\pi$). Then an argument similar to (viii-1) shows that $z$ is an arbitrary
element of $\mathcal{T}$.

\end{document}